\documentclass[%
 reprint,
superscriptaddress,
 amsmath,amssymb,
 aps,prc
]{revtex4-1}

\usepackage{graphicx}
\usepackage{dcolumn}
\usepackage{bm}
\usepackage{color}
\usepackage{titlesec}

\definecolor{mygreen}{rgb}{0.1, 0.6, 0.1}

\begin{document}

\preprint{APS/123-QED}

\title{New mass anchor points for neutron-deficient heavy nuclei from direct mass measurements of radium and actinium isotopes}

\author{M. Rosenbusch}
 \email{marco.rosenbusch@riken.jp}
\affiliation{%
 RIKEN Nishina Center for Accelerator-Based Science, Wako, Saitama 351-0198, Japan\\
}%

\author{Y. Ito}%
\affiliation{%
 RIKEN Nishina Center for Accelerator-Based Science, Wako, Saitama 351-0198, Japan\\
}%
\author{P. Schury}%
\affiliation{%
 Wako Nuclear Science Center (WNSC), Institute of Particle and Nuclear Studies (IPNS), High Energy Accelerator Research Organization (KEK), Wako, Saitama 351-0198, Japan\\
}%
\author{M. Wada}%
\affiliation{%
 RIKEN Nishina Center for Accelerator-Based Science, Wako, Saitama 351-0198, Japan\\
}%
\affiliation{%
 Wako Nuclear Science Center (WNSC), Institute of Particle and Nuclear Studies (IPNS), High Energy Accelerator Research Organization (KEK), Wako, Saitama 351-0198, Japan\\
 }
\affiliation{%
 Institute of Physics, University of Tsukuba, Ibaraki 305-8571, Japan\\
}%
\author{D. Kaji}%
\affiliation{%
 RIKEN Nishina Center for Accelerator-Based Science, Wako, Saitama 351-0198, Japan\\
}%
\author{K. Morimoto}%
\affiliation{%
 RIKEN Nishina Center for Accelerator-Based Science, Wako, Saitama 351-0198, Japan\\
}%
\author{H. Haba}%
\affiliation{%
 RIKEN Nishina Center for Accelerator-Based Science, Wako, Saitama 351-0198, Japan\\
}%
\author{S. Kimura}%
\affiliation{%
 RIKEN Nishina Center for Accelerator-Based Science, Wako, Saitama 351-0198, Japan\\
}%
\affiliation{%
 Wako Nuclear Science Center (WNSC), Institute of Particle and Nuclear Studies (IPNS), High Energy Accelerator Research Organization (KEK), Wako, Saitama 351-0198, Japan\\
 }
\affiliation{%
 Institute of Physics, University of Tsukuba, Ibaraki 305-8571, Japan\\
}%
\author{H. Koura}%
\affiliation{%
 Advanced Science Research Center, Japan Atomic Energy Agency, Ibaraki 319-1195, Japan\\
}%
\author{M. MacCormick}%
\affiliation{%
 Institut de Physique Nucl\'eaire, IN2P3-CNRS, Universit\'e Paris-Sud, Universit\'e Paris-Saclay, 91406 Orsay Cedex, France\\
}%
\author{H. Miyatake}%
\affiliation{%
 Wako Nuclear Science Center (WNSC), Institute of Particle and Nuclear Studies (IPNS), High Energy Accelerator Research Organization (KEK), Wako, Saitama 351-0198, Japan\\
}%
\author{J. Y. Moon}%
\affiliation{%
 Institute for Basic Science, 70, Yuseong-daero 1689-gil, Yusung-gu, Daejeon 305-811, Korea\\
}%
\author{K. Morita}%
\affiliation{%
 RIKEN Nishina Center for Accelerator-Based Science, Wako, Saitama 351-0198, Japan\\
}%
\affiliation{%
 Kyushu University, Nishi-ku, Fukuoka 819-0395, Japan
}%
\author{I. Murray}%
\affiliation{%
 Institut de Physique Nucl\'eaire, IN2P3-CNRS, Universit\'e Paris-Sud, Universit\'e Paris-Saclay, 91406 Orsay Cedex, France\\
}%
\author{T. Niwase}%
\affiliation{%
 RIKEN Nishina Center for Accelerator-Based Science, Wako, Saitama 351-0198, Japan\\
}%
\affiliation{%
 Kyushu University, Nishi-ku, Fukuoka 819-0395, Japan
}%
\author{A. Ozawa}%
\affiliation{%
 Institute of Physics, University of Tsukuba, Ibaraki 305-8571, Japan\\
}%
\author{M. Reponen}%
\affiliation{%
 RIKEN Nishina Center for Accelerator-Based Science, Wako, Saitama 351-0198, Japan\\
}%
\author{A. Takamine}%
\affiliation{%
 RIKEN Nishina Center for Accelerator-Based Science, Wako, Saitama 351-0198, Japan\\
}%
\author{T. Tanaka}%
\affiliation{%
 RIKEN Nishina Center for Accelerator-Based Science, Wako, Saitama 351-0198, Japan\\
}%
\affiliation{%
 Kyushu University, Nishi-ku, Fukuoka 819-0395, Japan
}%
\author{H. Wollnik}%
\affiliation{%
 New Mexico State University, Las Cruces, NM 88001, USA\\
}%

\date{\today}

\begin{abstract}
The masses of the exotic isotopes $^{210-214}$Ac and $^{210-214}$Ra have been measured with a multi-reflection time-of-flight mass spectrograph. These isotopes were obtained in flight as fusion-evaporation products behind the gas-filled recoil ion separator GARIS-II at RIKEN. The new direct mass measurements serve as an independent and direct benchmark for existing $\alpha$-$\gamma$ spectroscopy data in this mass region. Further, new mass anchor points are set for U and Np nuclei close to the $N=126$ shell closure for a future benchmark of the $Z=92$ sub-shell for neutron-deficient heavy isotopes. Our mass results are in general in good agreement with the previously indirectly-determined mass values. Together with the measurement data, reasons for possible mass ambiguities from decay-data links between ground states are discussed.
\end{abstract}

\maketitle


\section{\label{Intro}Introduction}

Heavy, neutron-deficient nuclei are highly interesting for studying the robustness of the $N=126$ shell closure and also the yet unresolved existence of a proton sub-shell at $Z=92$ for nuclides with less than $134$ neutrons. Due to the dominance of $\alpha$-decay in this region, the primary method applied to the investigation of shell structure is the measurement of decay radiation providing reduced decay widths \cite[]{Andreyev2013}. The $N=126$ and $Z=82$ double shell closure is the initiator of a region of extremely short-lived nuclei, where some aspects, such as the decay of $^{212}$Po, are already well understood \cite[]{Betan2012}, but properties of many other nuclei along the neutron shell are challenging for theory (see, {\emph e.g.}, recent laser spectroscopy of $^{214}$Fr and the discussion therein \cite[]{Flaroog2016}). Towards heavier isotopes the existence of a $Z=92$ proton shell closure near $N=126$ is discussed in literature but the experimental information is still scarce. First light could be shed via the decay of uranium and neptunium nuclei around $N=126$ \cite[]{Andreyev1992,Andreyev1993,Lepanen2007,Khuyagbaatar2015,Yang2015,SUN2017303}, indicating its non-existence in disagreement with state-of-the-art nuclear models (as discussed in \cite[]{Lepanen2007,Khuyagbaatar2015,SUN2017303}).\par%
The nuclear mass, as a direct observable for the binding energy, is another major ingredient for the study of nuclear shell evolution and provides independent signatures obtained from particle-separation energies \cite[]{Lunney2003}. The masses of large fraction of heavy nuclei have so far been determined via total decay energies ($Q_\alpha$ values). To do this, an accurate measurement of the decay-$\alpha$ kinetic energy, adjusted for recoil corrections is required (see \cite[]{SF2017} for the discussions).\par
Experimentally determined $Q_\alpha$ values allow to directly connect nuclear ground-state masses to the most precisely known mass in a given $\alpha$-decay chain. If direct transitions between ground states are not available, a combination of internal transition energies with intermediate decay-energies is necessary to reconstruct the full energetics of the path to the final state \cite[]{AME2016}. For the latter case, an unambiguous knowledge of spin, parity, and all transition energies is necessary. This necessity is a general feature for the use of spectroscopy or reaction data. For the detection of subsequent $\gamma$-rays after the decay, an important issue is also the conversion of the outgoing de-excitation photon into an electron from the atomic shell. While low-energy nuclear transitions are known to be almost fully converted, transition photons with energies above $100\,\mathrm{keV}$ may also be resonantly converted and are potentially not detected in $\alpha$-$\gamma$ measurements. Regarding the small average distance ($<300\,\mathrm{keV}$) of two-neutron separation energies for $N=126$ isotones above $^{206}$Pb, mass corrections at the order of $100\,\mathrm{keV}$ can contain significant nuclear-structure information.\par%
Prominent evidence for the risks of indirect mass measurements is the case of $^{150}$Ho, where the isomeric state was misidentified as being the ground state using $\beta$-endpoint spectroscopy \cite[]{Alkhazov1983}, as revealed through an $800\,\mathrm{keV}$ offset from Penning-trap direct mass measurements at CERN/ISOLDE \cite{Lunney2003,Beck2000a}. Experiments combining direct mass measurement with simultaneous spectroscopic observations allowed to determine the mass and spin-parities of the $^{190}$Tl $2^{-}$ ground and $7^{+}$ isomer state, and subsequently a clarification of ten mass values from two separate $\alpha$ chains has been obtained \cite[]{Stanja2013}. Very recently, another six isotope masses were disentangled by the mass and spin-parity assignments ($3/2^{-}$ and $13/2^{+}$ state) of $^{195,197}$Po \cite[]{Numa2017}. These studies show that for cases where an assignment is questionable, direct mass measurements are the only unambiguous access to the correct mass value.\par
A powerful method for high-precision direct mass measurements is multi-reflection time-of-flight mass spectrometry \cite[]{WOLLNIK1990267}. The experiments described here were carried out at RIKEN with the multi-reflection time-of-flight mass spectrograph (MRTOF-MS), which allows for simultaneous mass measurements of multiple isobaric chains \cite[]{SCHURY201419,SCHURY201439,SCHURY2016425,SCHURY2017a,Ito2017}. The reported measurements probe nuclei which are currently only indirectly determined, with the aim of consolidating the presently adopted mass values and to remove eventual ambiguities. In the following, the experimental setup and detailed data-analysis method will be discussed, and the results are compared with the current atomic-mass evaluation (AME2016) \cite[]{AME2016}.\\%

\section{Experiment}
\label{sec:Exp}

The nuclei $^{210-214}$Ra and $^{210-214}$Ac have been produced by bombarding multiple thin $^{169}$Tm targets with a high-intensity $^{48}$Ca beam. Sixteen $\sim$1~mg/cm$^2$ targets, were mounted on a rotatable wheel \cite[]{KAJI2008198} to accommodate a high beam intensity ($\gtrsim 1\,\mathrm{p\mu A}$), delivered by the RILAC accelerator at RIKEN \cite[]{RILAC1984}. A variety of evaporation channels were populated through the variable projectile energy between $5\,\mathrm{MeV/A}$ and $6\,\mathrm{MeV/A}$. The evaporation channels to actinium and radium isotopes were $^{169}$Tm($^{48}$Ca,xn)$^{217-x}$Ac (evaporation of $x$ neutrons) and $^{169}$Tm($^{48}$Ca,pxn)$^{216-x}$Ra (a proton and $x$ neutrons), respectively. Evaporation residues (EvR) were separated from the primary beam in the gas-filled recoil ion separator GARIS-II \cite[]{KAJI2013311}.\par
\begin{figure}[b]
  \includegraphics[width=0.48\textwidth]{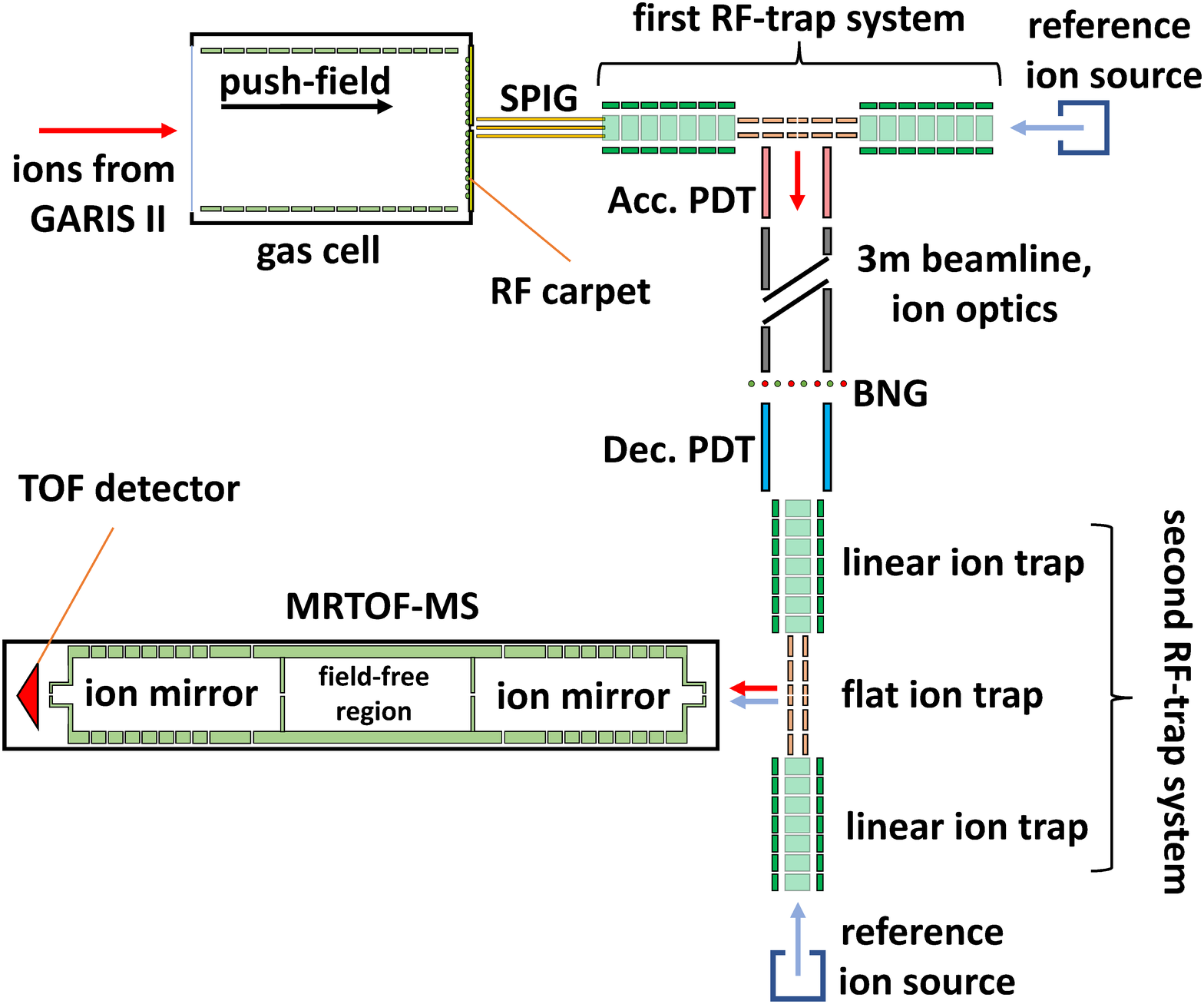}
  \caption{\label{fig:setup} Sketch of the SHE mass facility at RIKEN. The setup consists of a gas-filled stopping cell, two RFQ ion-trap systems including each a flat ion trap, a transport section with a Bradbury-Nielsen gate for mass separation, and a multi-reflection time-of-flight mass spectrograph.}
\end{figure}

%
EvR masses were determined using the super-heavy element (SHE) mass facility of RIKEN-KEK \cite[]{SCHURY201439,SCHURY2016425}. The long-term goal of this facility is the identification of newly synthesized super-heavy elements by their mass value, and to provide new, accurate values with a relative uncertainty of $\delta m / m < 10^{-6}$. A schematic sketch of the setup is shown in Fig.~\ref{fig:setup}.\par%
To successfully measure EvR's with the MRTOF-MS, the initial kinetic energies must be reduced to a value suitable for trapped-ion mass measurements. A helium-gas filled stopping cell ($p=10^{4}\,\mathrm{Pa}$) and additional beam-energy degraders (Mylar foils) of adjustable thickness of $10-13\,\mathrm{\mu m}$ have been used to slow down and stop EvR ions inside the gas cell. Cryogenic temperatures were used to freeze-out volatile stable contaminants and EvR ions were thermalized in stochastic processes with the gas atoms. Extracted ions were seen to be singly or doubly charged, with an unexpected dominant component of doubly charged ions \cite[]{SCHURY2017160}. Cylindrical guiding electrodes in the gas cell push the ions to the surface of a circular-shaped radio-frequency (RF) ion carpet \cite[]{WADA2003570,BOLLEN2011131} operated at roughly $8\,\mathrm{MHz}$ and $\approx150\,\mathrm{Vpp}$ amplitude, to prevent the ions from hitting the cavity surface. The ions are then guided along concentric RF electrode rings to a small central exit hole using the traveling-wave technique \cite[]{ARAI201456} and are extracted by the electric DC field.\par%
An RF sextupole ion guide (SPIG) \cite[]{XU1993274} with an inner radius of about $2\,\mathrm{mm}$ transports the ions to a system consisting of three linear radio-frequency quadrupole (RFQ) ion traps. The central trap (referred to as ``flat ion trap'') has a planar electrode design printed on circuit board \cite[]{Schury2009}, which allows to eject ions orthogonally in a high-quality dipole field. The orthogonal ejection further allows to accept ions from the linear RFQ traps on either side (one injection at a time in present operation), which after this first trap system is used for independent beam tuning. After pre-cooling in a helium environment ($p=1\,\mathrm{Pa}$) in the outer RFQ, the ions are transmitted to the flat ion trap, where the ion cloud diameter is reduced to less than $300\,\mathrm{\mu m}$ by further cooling as preparation for the extraction of ions. Using this symmetric trap design, both the transmission of the analyte ions from the gas cell and the transmission of reference ions coming from an alkali-ion source towards the mass-measurement section is realized under the same conditions.\par%
After extraction from the flat ion trap, the ions are accelerated by a pulse drift tube ("Acc. PDT" in Fig.~\ref{fig:setup}) to an average kinetic energy of $1\,\mathrm{keV}$ towards the mass-measurement section. Between the gas cell and mass measurement section, the ions travel through about $3\,\mathrm{m}$ of beam line. A Bradbury-Nielsen ion gate (BNG) \cite[]{Bradbury1936} has been installed for mass selection by the ions' time of flight after ejection from the trap, where a resolving power of $R\approx100$ can be reached. Further downstream, the mass-selected ions are decelerated prior to another RFQ ion-trap system of identical construction to the one discussed above. The deceleration process after transfer takes place using a second pulsed drift tube (``Dec. PDT'') which is placed in a region where the helium pressure originating from the second RFQ trap system is already sufficient to ensure atom-ion collisions during the slowing process. In this way, molecules are broken up, further reducing contamination by stable molecular ions.\par%
      \begin{figure}[t]
	\centering
	\includegraphics[width=1.0\linewidth]{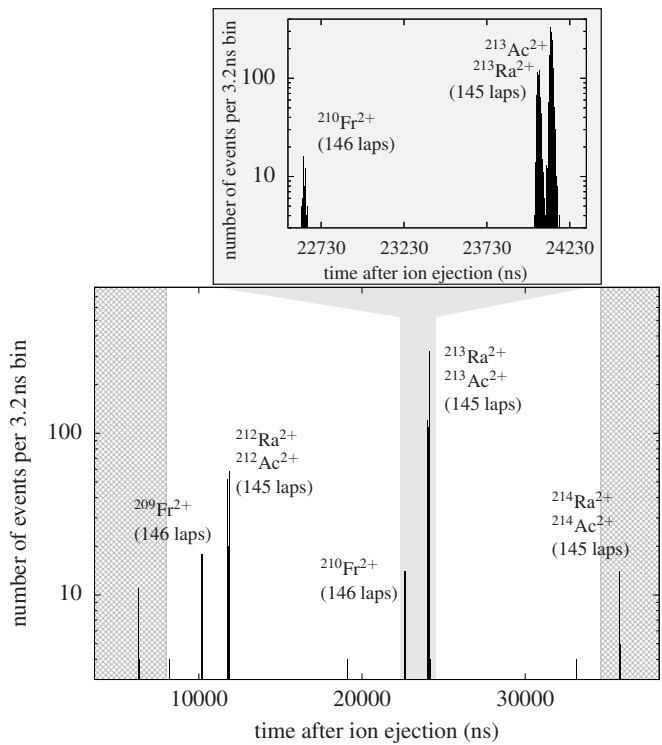}
	\caption{Ion count as a function of the arrival time after the ejection from the MRTOF-MS. The spectrum contains several ion species which are labeled at each signal peak. The hatched areas denote the time regions in which the ion peaks are not used for precision measurements due to the voltage transition at the ejection-mirror (see text). Background counts and further ion peaks of up to 3 counts are not shown for a better overview. The small figure on top shows an enlargement of the grey shaded region in the lower figure to illustrate the isobaric separation of $^{213}$Ac and $^{213}$Ra.}
	\label{fig:wideband}
      \end{figure}
After deceleration, the analyte ions are captured in the upstream RFQ part of the second ion-trap system to be transferred to the flat ion trap. In the downstream RFQ trap, reference ions are prepared at the same time to enable an alternate cycle-by-cycle ejection of either species for interleaved measurement of analyte ions and reference ions. This method has been called ``concomitant referencing'' and was demonstrated recently \cite[]{SCHURY2017a}. One of the major features enabled by the concomitant referencing method is a treatment of the spectra of the analyte ions and the reference spectra as one data set allowing for correction of time-of-flight drifts within the MRTOF-MS (see Section~\ref{sec:analysis})~\cite[]{SCHURY201439}.\par%
The analyte ions are orthogonally ejected from the flat ion trap and transported to the MRTOF-MS, which consists of two concentric ion mirrors separated by a field-free region enabling a closed ion trajectory by multiple ion reflections. Capturing of ions is performed by lowering (and re-raising) the electric potential of the first mirror electrode on the injection side, while ejection is accomplished by lowering the potential of the conjugate mirror electrode on the ejection side. The ions are injected and stored for a defined duration causing the ions to perform multiple laps (full motional oscillations between the two ion mirrors) and are ejected towards a fast TOF detector (``MagneTOF'') which is mounted next to the mirror on the ejection side. The signals from the ion impacts are forwarded to a time-to-digital converter (TDC) model MCS6A from FAST ComTec, using the trigger for ion ejection from the flat ion trap as start signal. The time resolution of the TDC is $100\,\mathrm{ps}$. The masses of the ions are obtained from the TOF as described in Section~\ref{sec:analysis}.\par%
The number of laps has been varied ($144,145$, and $146\,\mathrm{laps}$) to exclude misinterpretations of ion signals from species of significantly different charge-to-mass ratios performing a different number of laps than the analyte ions. Such ions can appear in the same spectrum due to the periodicity of the system. If the lap time is precisely adapted to the analyte ions (and similar masses, such as those of isobars), an intruder signal of different charge-to-mass ratio is identified by its change in relative position in the spectrum when the number of laps is altered, whereas for ions with correct circulation times the shift of the signal positions is typically not visible \cite[]{SCHURY201419}. The cycle time of the mass measurement including both analyte ions and reference measurement, has a total period of $30\,\mathrm{ms}$ (repetition frequency of $33\,\mathrm{Hz}$).\par%
A time-of-flight spectrum including the maximum meaningful ejection-time window for doubly-charged $A/q~\sim~100$ ions is shown in Fig.~\ref{fig:wideband}. Due to the necessity of changing electric potentials for the ion ejection, ions detected at the left and right side of the spectrum (hatched regions) are close to the ejection mirror at the moment of voltage switching and are thus disturbed in their motion. As this results in deviant mass determination, these time regions are unused. In the allowed time window, for $A/q \approx 100$, ions spanning a range of just over two mass units will make the same number of laps. Other species with different numbers of laps can be identified within the spectra. The inset of Fig.~\ref{fig:wideband} shows a magnification where isobaric separation is visible, even in logarithmic scale. The mass resolving power achieved in this experiment was $R \approx 130000$.\par%
\section{Data Analysis}
\label{sec:analysis}
The time-of-flight data has been analyzed using the binned maximum-likelihood estimator provided by the ROOT package \cite[]{BRUN199781} developed at CERN. As ion peaks from MR-TOF spectra often have non-Gaussian shapes, meanwhile asymmetric fit functions are preferred to provide higher accuracies as, \emph{e.g.}, reported in a recent analysis of transfermium nuclei \cite[]{Ito2017}. The chosen probability density function (PDF) in the present work is a Gaussian hybrid with two exponential tails previously applied for $\gamma$-spectroscopy \cite[]{KOSKELO198111}. To provide an overview, the function definition that exists in three different intervals is shown in Eq.~\ref{eq:Fitfunction}, where $A$ is the amplitude, the running variable $t$ is the time after the start of the TDC, $t_\mathrm{c}$ the position of the maximum of the function interpreted as the central detection time, and $w$ is the width of the Gaussian fit function.\par
On both sides of the Gaussian shape, the function changes smoothly into exponential curves at the transition points, $t_\mathrm{L}= t_c - \delta_\mathrm{L}$ on the left side (smaller $t$), and $t_\mathrm{R} = t_c + \delta_\mathrm{R}$ on the right side (larger $t$), being defined such that the function and its derivative are continuous. For the signal shapes of time-of-flight data, no satisfactory theoretical models are presently available. However, a deviation from a Gaussian shape at arbitrary distances from the peak center is a common feature for MRTOF-MS spectra, a feature which is encountered also in other spectroscopic fields. Previously, a single-sided exponential-Gaussian hybrid function used for chromatography \cite[]{LAN20011} yielded improvements over a Gaussian fitting function for MRTOF-MS spectra \cite[]{SCHURY201439,ITO2013}. The presently used Gaussian hybrid with an additional degree of freedom yields better phenomenological agreement. A high tolerance for asymmetries is achieved and their bias on the extracted center value is satisfyingly low.\\%
\begin{figure}[t]
  \includegraphics[width=0.45\textwidth]{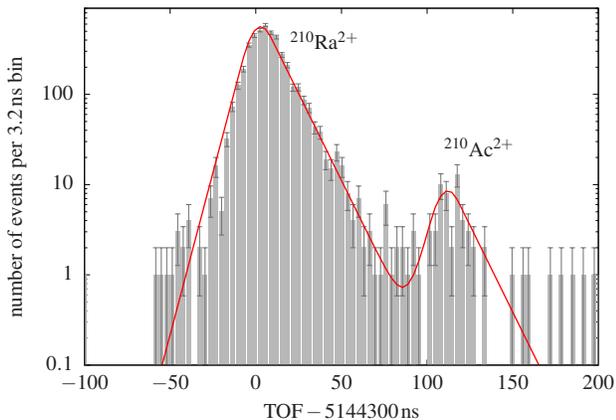}
  \caption{\label{fig:fitplot} Ion count as a function of the detection time after start of the TDC. Red line: Gaussian hybrid fit function for two adjacent isobaric-ion peaks.}
\end{figure}
%
%
%
  \begin{eqnarray}
    \label{eq:Fitfunction}
    &f(t) &{}=A\cdot
    \begin{cases}
      \mathrm{exp}\left( \frac{\delta_\mathrm{L} \left(  2t - 2t_\mathrm{c} + \delta_\mathrm{L}\right)}{2w^2} \right) & t \leq t_\mathrm{L} \\
	    \mathrm{exp}\left( \frac{-\left(t - t_\mathrm{c}\right)^2}{2w^2}\right)			  &t_\mathrm{L} < t < t_\mathrm{R} \\
	    \mathrm{exp}\left( \frac{\delta_\mathrm{R} \left(- 2t + 2t_\mathrm{c} + \delta_\mathrm{R} \right)}{2w^2} \right)	  &t \geq t_\mathrm{R}\quad\mathrm{.} \\
    \end{cases}
  \end{eqnarray}
  %
%

In general, when analyzing isobar chains, a sum of peaks as described in Eq.~\ref{eq:Fitfunction} is employed as a merged PDF with shared values of $\delta_\mathrm{L}$, $\delta_\mathrm{R}$, and $w$. An example of such fitting for an isobaric doublet, is shown in Fig.~\ref{fig:fitplot}. The center values of the detection time and their uncertainties (same for other parameters) are provided by the MINUIT package of ROOT.\par
The correction of TOF drifts during the measurement time is performed by dividing the data in subsets of certain intervals of the measurement time. In every subset $j$ the central detection time of the reference ions $t_{\mathrm{c},j}^\mathrm{ref}$ is extracted from the fit to the data of the subset, and then the subsets are summed again after modifying the time of each single ion event $t_{i,j}^\mathrm{raw}$, where $i$ spans over all ions (reference and analyte). As the presumed sources of TOF drift -- voltage fluctuations and thermal expansions -- are of ratiometric nature, a multiplication of the ratio $t_{\mathrm{c},j}^\mathrm{ref} / \left<t_\mathrm{c}^\mathrm{ref} \right>$ is performed to correct for the drifts, where $\left<t_\mathrm{c}^\mathrm{ref} \right>$ is the average value of all fitted reference times:
\begin{equation}
  \label{eq:correct}
  t_{i, j}^\mathrm{correct}= \left.t_{i,j}^{\mathrm{raw}}\cdot \frac{\left< t_\mathrm{c}^\mathrm{ref} \right>}{t_{\mathrm{c},j}^\mathrm{ref}}\right..
\end{equation}
If the time of flight $t_\mathrm{c}$ of the analyte ion is known, the mass $m$ of this ion with charge $q$ can be calculated by:
\begin{equation}
  \label{eq:mass}
    m = \frac{q}{q_\mathrm{ref}} m_\mathrm{ref} \cdot \rho^2\quad.
\end{equation}
Here, $\rho = (t_\mathrm{c} - t_0)\,/\,(t_\mathrm{c}^\mathrm{ref} - t_0)$ is the ratio of the effective TOF of the analyte ion and that of the reference ion, and $m_\mathrm{ref}$ and $q_\mathrm{ref}$ are the mass and the charge of the reference ion, respectively. The offset time $t_0$ is the time difference of the ejection pulse of the flat ion trap to the start of the TDC.\par%
One way to determine $t_0$ is to make use of a second ion species of well-known mass to eliminate the $t_0$ parameter (see {\emph e.g.} \cite[]{Wienholtz2013}). However, two simultaneously measured reference masses are not always available in the same spectrum. For the SHE mass facility an alternate method has been chosen: The delay $t_0$ of the electronic signals between the ejection pulse of the flat ion trap and the start trigger of the TDC has been measured directly. Furthermore, the upper and the lower limits of $t_0$ have been estimated conservatively covering uncertainties in cable lengths. This measurement yields $t_0 = 45(5)\,\mathrm{ns}$ where the uncertainty of $t_0$ yields a systematic mass uncertainty. This calibration and the concomitant referencing method enables precision mass measurements independent of well-known species produced online.\par
In an early use of single-reference mass measurements for $^{8}$Li$^{+}$ ions using $^{12}$C$^{+}$ as reference ions \cite[]{ITO2013} (with a previous setup) an assumed uncertainty of $\delta t_0 = 10\,\mathrm{ns}$ resulted in a systematic mass uncertainty of $3.4\,\mathrm{keV}$ (relative contribution $4.5\times 10^{-7}$). This was achieved by a long flight path $n>500\,\mathrm{laps}$. However, the present setup is tuned for fast measurements $n<200\,\mathrm{laps}$, where the result of this study is more affected. The relative mass uncertainty further scales with the charge state of the ions (doubly-charged $A~\sim~210$ ions and singly-charged $^{133}$Cs$^{+}$ reference), where an uncertainty of $\delta t_0 = 5\,\mathrm{ns}$ yields a systematic mass uncertainty of the order of $\delta m_\mathrm{sys} = 40\,\mathrm{keV}$ for the present mass region.\par
\begin{figure}[t]
  \centering
  \includegraphics[width=0.9\linewidth]{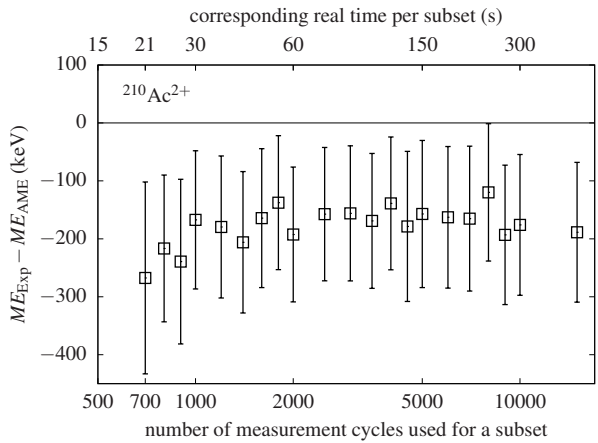}
  \caption{Difference of the measured mass excess of $^{210}$Ac to the value listed in AME2016 as a function of the number of measurement cycles used for each subset to perform the TOF drift correction.}
  \label{fig:scatter_test}
\end{figure}
To reduce the systematic mass uncertainty, $^{213}$Ra as measured with high statistics in the present experiment, and as also known from high-precision, high-reliability Penning-trap data \cite[]{Droese2013}, has been used to provide a more accurate calibration of $t_0$. For small offset times as compared to the total flight time (typically $t_0/t_c \approx 10^{-5}$) the dependency of the mass on $t_0$ is linear, which can be seen from the first order expansion with respect to $t_0$ in Eq.~\ref{eq:linear} \cite[]{ITO2013}. 
\begin{equation}
  \label{eq:linear}
  m=\frac{q}{q_\mathrm{ref}}\left(m_\mathrm{ref}\left(\frac{t_\mathrm{c}}{t_\mathrm{c}^\mathrm{ref}}\right) + 2m_\mathrm{ref} \frac{t_\mathrm{c}\left(t_\mathrm{c}-t_\mathrm{c}^\mathrm{ref}\right)}{\left(t_\mathrm{c}^\mathrm{ref}\right)^3}t_0\right)
\end{equation}
The mass increase as a function of $t_0$ has been determined using two different values of $t_0$ for $^{213}$Ra in Eq.~\ref{eq:mass}, yielding a coefficient of $\Delta m(\Delta t_0)/\Delta t_0 = 7.98\,\mathrm{keV/ns}$. The Birge-ratio adjusted uncertainty of $8.01\,\mathrm{keV}$ for $^{213}$Ra and its uncertainty listed in AME2016 \cite[]{AME2016} of $9.82\,\mathrm{keV}$ yield a combined systematic uncertainty of $\delta m_\mathrm{sys} = 12.67\,\mathrm{keV}$. Adjusting our value to the AME2016 value, the effective offset time was determined to be $t_0=52.72(1.59)\,\mathrm{ns}$, where the uncertainty for $t_0$ is obtained from the values of $t_0$ at which $m_\mathrm{Exp} - m_\mathrm{AME} = \delta m_\mathrm{sys}$. This was used to calculate all other masses (systematic uncertainties are indicated in Table~\ref{tab:mass_values}). As can be seen from the dependency on $t_c$ in Eq.~\ref{eq:linear} with $t_c^\mathrm{ref}$ being constant, the systematic uncertainty $\delta m_\mathrm{sys}$ scales slightly with the mass number.\par
This calibration is of hybrid type using one well-known species in the region as a second mass reference for the calibration of $t_0$, but using the single-referencing method with $^{133}$Cs$^{+}$ ions otherwise. One consideration is that $^{213}$Ra possesses an isomer with an excitation energy of $1768(4)\,\mathrm{keV}$ and a half life of $2.2\,\mathrm{ms}$. However, the travel time inside the gas cell is expected to be $30\,\mathrm{ms}$ in average. Including further durations until the ion detection in the present case of operation, the isomer is suppressed by a factor of about $10^5$.\par
During the data analysis an additional study has been performed in order to take into account eventual shifts of the measured values $t_\mathrm{c}$ induced by the boundaries chosen for the data fits. One of the most important degrees of freedom is the length of the time intervals ({\emph i.e.}, number of measurement cycles) chosen for each data subset to perform the drift correction. According to the detection of maximum 30 reference ions per second (double counts of reference ions are presently avoided), TOF drift durations on the order of a few of seconds can be resolved, dependent on the required precision. If the maximum drift frequency is not fully resolved, random fluctuations of the mass results can be observed when the time intervals are varied, but the nature of these is not necessarily stochastic. A wrapping program for the ROOT fitting routine has been developed which enables multiple fits using various different boundaries automatically. An example of the mass scattering for different numbers of measurement cycles for each subset used for drift correction is shown in Fig.~\ref{fig:scatter_test}.\par%
Additionally, the effect of choice of fitting range, bin size, and initial bin origin have been investigated. The result of this study yields an additional average scattering of the results within $30-40\,\mathrm{\%}$ of the statistical uncertainty. Values from failed fits, or those significantly away from the mean value, have been excluded. For the two cases with lowest statistics, \emph{i.e.}, $^{210,211}$Ac next to the much more abundant $^{210,211}$Ra peaks (see Fig.~\ref{fig:fitplot}), this study has revealed enhanced sensitivities of the mass result, which may suggest to include an additional mass uncertainty. In the investigated parameter set, a standard deviation of the scattering of $1.45\,\sigma$ for $^{210}$Ac and $1.65\,\sigma$ for $^{211}$Ac was observed. Since, additionally, no Birge ratio is available for these data sets, these numbers have been included for the suggested mass uncertainties in Table~\ref{tab:mass_values}.\\%
\section{Results and Discussion}
\label{results}
\begin{figure}[t]
\includegraphics[width=0.48\textwidth]{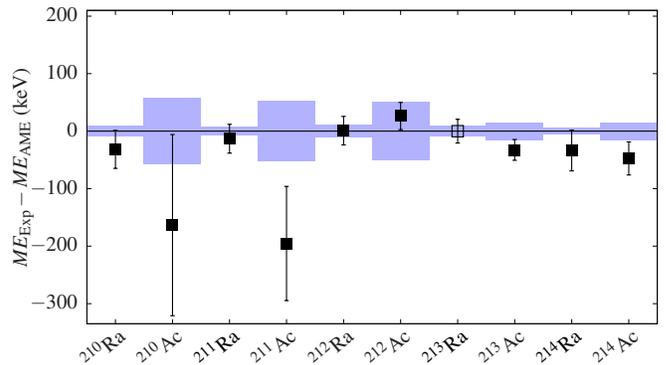}
\caption{\label{fig:results} Difference of the present masses from those listed in AME2016. The blue shadows denote the uncertainties from AME2016. The open symbol and zero distance from the AME2016 value for $^{213}$Ra indicates that this mass has been used for calibration.}
\end{figure}

The resulting mass values of this measurement are listed in Table~\ref{tab:mass_values}. As the given raw data refers to doubly-charged ions measured with $^{133}$Cs$^{+}$ ions as reference, the atomic masses $m$ are derived from the squared TOF ratios by:
\begin{equation}
 m = \frac{q}{q_\mathrm{ref}}(m(^{133}\mathrm{Cs}) - m_e)\cdot \rho^2 + 2m_e ,
\end{equation}
  \begin{table*}
	  \caption{Measured isotopes, square of time-of-flight ratios $\rho^2$, Birge-ratio $BR$ of the data sets (if available, 1.2 otherwise), measured mass excess $ME$, mass excess from literature $ME_\mathrm{Lit}$ (AME2016), and difference of experimental value and literature value $\Delta ME = ME_\mathrm{Exp} - ME_\mathrm{Lit}$. For the $\rho^2$ values, the first bracket refers to the statistical uncertainty and the second bracket to the systematic contribution. Mass-excess uncertainties are derived using the statistical uncertainty corrected for the Birge ratio if availabe and larger than 1. If not available, the value 1.2 from $^{213}$Ac has been used. In the two cases where the evaluation results revealed additional sensitivity (marked by an asterisk), the mass uncertainty has been modified as discussed in Section.~\ref{sec:analysis}. The systematic contribution has been added to the scaled statistical uncertainty. For the AME input values ($\rho^2$), further digits than significant have been added.\\
      \\}%
	   
   \begin{tabular}{ c | c | c | c | c | c }
     \label{tab:mass_values}
     \quad species \quad  & \quad $\rho^2$ \quad & \quad $BR$ \quad & \quad $ME$ (keV) \quad & \quad $ME_\mathrm{Lit}$ (keV) \quad & \quad $\Delta ME$ (keV) \quad \\[1pt]
     \hline
     &&&&&\\[0.2pt]
           $^{210}$Ra$^{2+}$ &  0.7900361042(664)(540)    &	N/A (1.2)     	&	411.1	(33.1)	&	442.8(9.2)		  &-31.7	(34.4)	\\[3pt]
           $^{210}$Ac$^{2+}$ &  0.790069283(402)(54)	    &	*N/A (1.45)	&	8626	(157)	&	8790.7(57.4)		  &-163		(169)	\\[3pt]
           $^{211}$Ra$^{2+}$ &  0.7937998391(404)(530)    &	N/A (1.2)	&	818.8	(25.1)	&	832.0(7.9)		  &-13.2	(26.3)	\\[3pt]
           $^{211}$Ac$^{2+}$ &  0.793824831(211)(53)      &	*N/A (1.65)	&	7007	(99)	&	7202.2(53.0)		  &-195		(113)	\\[3pt]
	   $^{212}$Ra$^{2+}$ &  0.7975578207(479)(521)    &	0.94		&	-198.1	(24.8)	&	-199.0(11.3)		  &0.9		(27.2)	\\[3pt]
           $^{212}$Ac$^{2+}$ &  0.7975881181(409)(521)    &	1.07		&	7303.6	(23.7)	&	7277.3(51.4)		  &26.3		(56.6)	\\[3pt]
           $^{213}$Ac$^{2+}$ &  0.8013454345(181)(512)    &	1.20		&	6122.0	(18.1)	&	6154.7(15.3)		  &-32.7	(23.7)	\\[3pt]
	   $^{214}$Ra$^{2+}$ &  0.8050830358(929)(503)    &	0.65		&	59.0	(35.4)	&	92.7(5.3)		  &-33.7	(35.8)	\\[3pt]
	   $^{214}$Ac$^{2+}$ &  0.8051086313(661)(503)    &	0.57		&	6396.4	(28.8)	&	6443.9(15.4)		  &-47.4	(32.7)	\\[3pt]
   \end{tabular}
	
  \end{table*}
where $m_e$ is the electron mass, $q=2$ for doubly-charged analyte ions, and $q_\mathrm{ref}=1$ for $^{133}$Cs$^{+}$. A graphical comparison of the present measurements and the values listed in AME2016 is shown in Fig.~\ref{fig:results}. The different radioactive species were produced simultaneously and about 50 data sets were recorded over the course of a day. Due to the particularly low yield of $^{210,211}$Ac, subsequently recorded data sets for the same number of laps (containing both $^{210,211}$Ac and $^{210,211}$Ra as in Fig.~\ref{fig:fitplot}) have been combined.\par%
Including the correction for the Birge ratio, the mean statistical relative uncertainty of the measurements is $\delta m / m = 1.9\times 10^{-7}$. The use of $^{213}$Ra as an additional reference in order to reduce the uncertainty of the offset time $t_0$ allowed to achieve an average combined (statistical and systematic) relative uncertainty of $2.5\times 10^{-7}$. In order to visualize the measurement precision in comparison to the other masses, $^{213}$Ra is shown with open symbol in Fig.~\ref{fig:results}.\par%
In general a good agreement with the existing data from AME2016 is observed, confirming the previously performed measurements, mainly via $\alpha$-$\gamma$ spectroscopy \cite[]{2000He17}. The two newly measured even-even isotopes $^{210,212}$Ra agree very well with the previous indirectly determined values. For those isotopes the situation for $\alpha$ transitions is very clear, and additional information from Penning-trap measurements of $^{206}$Rn \cite[]{Droese2013}, the $\alpha$ daughter of $^{210}$Ra, is available. Our measurement of $^{211}$Ra is in similarly good agreement with previous Penning trap measurements \cite[]{Kowalska2009}.\par%
For the newly-measured actinium isotopes, the situation is different. For none of these isotopes the assignment of ground-state spin and parity is known for both mother and daughter nucleus at the same time. Furthermore, for $^{210-214}$Ac the experimental knowledge of excited states is still scarce. In such situations the existence of an $\alpha$ transition between yet unrecognized states cannot be excluded. However, for $^{210,212-214}$Ac, our measurements are in agreement with the previous mass assignments on the $1.5 \sigma$ level.\par
The isotope, $^{211}$Ac, is in agreement with the present AME2016 mass value only on the $2\sigma$ level. The mass value of $^{211}$Ac was indirectly determined most recently via $\alpha$-$\gamma$ spectroscopy \cite[]{2000He17} linked to the daughter $^{207}$Fr. The mass of $^{207}$Fr has been measured at CERN/ISOLDE with $^{133}$Cs as the reference mass \cite[]{2014Bo26}. Our new mass excess value is $ME=7007(99)\mathrm{keV}$, which deviates by $\Delta m=-195(113)\mathrm{keV}$ ($\approx1.7\sigma$) from the AME2016 value. As the mass excess is smaller than that of the previous measurement, an influence of long-lived excited states (presently unknown) may be possible. \par 
\begin{figure}[t]
\includegraphics[width=0.48\textwidth]{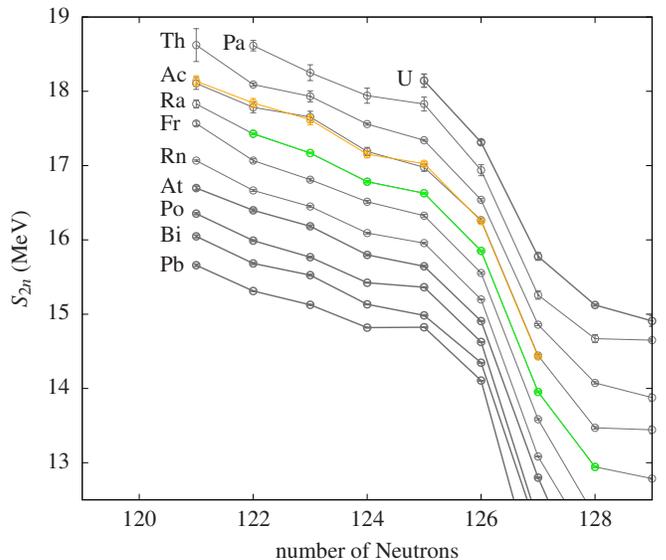}
\caption{\label{fig:S2N} Two-neutron separation energies for istoopic chains from Pb to U. The combined values of AME2016 and this work are indicated with color. Values from AME2016 are shown with grey lines.}
\end{figure}

To illustrate the two-neutron separation energies modified by the new direct mass measurements ($S_{2n}$), in Fig.~\ref{fig:S2N} we have combined our binding-energy values with the AME2016 values using the uncertainty-weighted mean value (no further weighting). The corresponding $S_{2n}$ values are indicated with color. As the masses of $^{210-214}$Ra are confirmed, no significant changes are found for the radium chain. For actinium isotopes, the two-neutron separation energies would mainly be influenced by the newly measured mass of $^{211}$Ac, but the impact is small due to the uncertainties, which exceed the presently adopted values by a factor of two. The combined values suggest minor changes as a small enhancement of the $N=125$ kink as seen in Pb, Po, At, and Pa isotopes and a smoothening of the trend at $N=122$ as opposed to the Th chain.%
\section{Summary and Conclusion}
\label{sec:summary_and_conclusion}
We have measured the masses of the isotopes $^{210-214}$Ra and $^{210-214}$Ac using multi-reflection time-of-flight mass spectrometry. All reported actinium isotopes and also $^{210,212}$Ra have been measured directly for the first time. The results are generally in good agreement with the previously listed values of the Atomic Mass Evaluation from 2016. For $^{211}$Ac a slight deviation of our measurements from the previous values have been obtained. This measurement proves once more the capability of multi-reflection devices to perform simultaneous mass measurements of several isotopes at the same time with isobaric resolution. The new direct measurements of Ac isotopes set anchor points for the decay chains of $^{218-222}$Np while the newly measured $^{210,212}$Ra isotopes do so for $^{218,220}$U. This work paves the way for further studies of the poorly investigated sub-shell closure at $Z=92$ near $N=126$ by reducing the length of decay chains for indirect mass determination to two decays for those nuclei.\\

\section{Acknowledgements}
\label{sec:acknowledgements}
We express our gratitude to the RIKEN Nishina Center for Accelerator-based Science and the Center for Nuclear Science at Tokyo University for their support of online measurements. We thank S. Naimi for helpful discussions about the previously measured data. This work was supported by the Japan Society for the Promotion of Science KAKENHI (Grants No. 2200823, No. 24224008, No. 24740142, No. 15H02096, No. 15K05116, 17H01081, and 17H06090).

\bibliography{Actinium_Radium}

\begin{thebibliography}{41}%
\makeatletter
\providecommand \@ifxundefined [1]{%
 \@ifx{#1\undefined}
}%
\providecommand \@ifnum [1]{%
 \ifnum #1\expandafter \@firstoftwo
 \else \expandafter \@secondoftwo
 \fi
}%
\providecommand \@ifx [1]{%
 \ifx #1\expandafter \@firstoftwo
 \else \expandafter \@secondoftwo
 \fi
}%
\providecommand \natexlab [1]{#1}%
\providecommand \enquote  [1]{``#1''}%
\providecommand \bibnamefont  [1]{#1}%
\providecommand \bibfnamefont [1]{#1}%
\providecommand \citenamefont [1]{#1}%
\providecommand \href@noop [0]{\@secondoftwo}%
\providecommand \href [0]{\begingroup \@sanitize@url \@href}%
\providecommand \@href[1]{\@@startlink{#1}\@@href}%
\providecommand \@@href[1]{\endgroup#1\@@endlink}%
\providecommand \@sanitize@url [0]{\catcode `\\12\catcode `\$12\catcode
  `\&12\catcode `\#12\catcode `\^12\catcode `\_12\catcode `\%12\relax}%
\providecommand \@@startlink[1]{}%
\providecommand \@@endlink[0]{}%
\providecommand \url  [0]{\begingroup\@sanitize@url \@url }%
\providecommand \@url [1]{\endgroup\@href {#1}{\urlprefix }}%
\providecommand \urlprefix  [0]{URL }%
\providecommand \Eprint [0]{\href }%
\providecommand \doibase [0]{http://dx.doi.org/}%
\providecommand \selectlanguage [0]{\@gobble}%
\providecommand \bibinfo  [0]{\@secondoftwo}%
\providecommand \bibfield  [0]{\@secondoftwo}%
\providecommand \translation [1]{[#1]}%
\providecommand \BibitemOpen [0]{}%
\providecommand \bibitemStop [0]{}%
\providecommand \bibitemNoStop [0]{.\EOS\space}%
\providecommand \EOS [0]{\spacefactor3000\relax}%
\providecommand \BibitemShut  [1]{\csname bibitem#1\endcsname}%
\let\auto@bib@innerbib\@empty
\bibitem [{\citenamefont {Andreyev}\ \emph {et~al.}(2013)\citenamefont
  {Andreyev}, \citenamefont {Huyse}, \citenamefont {Van~Duppen}, \citenamefont
  {Qi}, \citenamefont {Liotta}, \citenamefont {Antalic}, \citenamefont
  {Ackermann}, \citenamefont {Franchoo}, \citenamefont {He\ss{}berger},
  \citenamefont {Hofmann}, \citenamefont {Kojouharov}, \citenamefont {Kindler},
  \citenamefont {Kuusiniemi}, \citenamefont {Lesher}, \citenamefont {Lommel},
  \citenamefont {Mann}, \citenamefont {Nishio}, \citenamefont {Page},
  \citenamefont {Streicher}, \citenamefont {\ifmmode~\check{S}\else
  \v{S}\fi{}\'aro}, \citenamefont {Sulignano}, \citenamefont {Wiseman},\ and\
  \citenamefont {Wyss}}]{Andreyev2013}%
  \BibitemOpen
  \bibfield  {author} {\bibinfo {author} {\bibfnamefont {A.~N.}\ \bibnamefont
  {Andreyev}}, \bibinfo {author} {\bibfnamefont {M.}~\bibnamefont {Huyse}},
  \bibinfo {author} {\bibfnamefont {P.}~\bibnamefont {Van~Duppen}}, \bibinfo
  {author} {\bibfnamefont {C.}~\bibnamefont {Qi}}, \bibinfo {author}
  {\bibfnamefont {R.~J.}\ \bibnamefont {Liotta}}, \bibinfo {author}
  {\bibfnamefont {S.}~\bibnamefont {Antalic}}, \bibinfo {author} {\bibfnamefont
  {D.}~\bibnamefont {Ackermann}}, \bibinfo {author} {\bibfnamefont
  {S.}~\bibnamefont {Franchoo}}, \bibinfo {author} {\bibfnamefont {F.~P.}\
  \bibnamefont {He\ss{}berger}}, \bibinfo {author} {\bibfnamefont
  {S.}~\bibnamefont {Hofmann}}, \bibinfo {author} {\bibfnamefont
  {I.}~\bibnamefont {Kojouharov}}, \bibinfo {author} {\bibfnamefont
  {B.}~\bibnamefont {Kindler}}, \bibinfo {author} {\bibfnamefont
  {P.}~\bibnamefont {Kuusiniemi}}, \bibinfo {author} {\bibfnamefont {S.~R.}\
  \bibnamefont {Lesher}}, \bibinfo {author} {\bibfnamefont {B.}~\bibnamefont
  {Lommel}}, \bibinfo {author} {\bibfnamefont {R.}~\bibnamefont {Mann}},
  \bibinfo {author} {\bibfnamefont {K.}~\bibnamefont {Nishio}}, \bibinfo
  {author} {\bibfnamefont {R.~D.}\ \bibnamefont {Page}}, \bibinfo {author}
  {\bibfnamefont {B.}~\bibnamefont {Streicher}}, \bibinfo {author}
  {\bibfnamefont {i.~c.~v.}\ \bibnamefont {\ifmmode~\check{S}\else
  \v{S}\fi{}\'aro}}, \bibinfo {author} {\bibfnamefont {B.}~\bibnamefont
  {Sulignano}}, \bibinfo {author} {\bibfnamefont {D.}~\bibnamefont {Wiseman}},
  \ and\ \bibinfo {author} {\bibfnamefont {R.~A.}\ \bibnamefont {Wyss}},\
  }\href {\doibase 10.1103/PhysRevLett.110.242502} {\bibfield  {journal}
  {\bibinfo  {journal} {Phys. Rev. Lett.}\ }\textbf {\bibinfo {volume} {110}},\
  \bibinfo {pages} {242502} (\bibinfo {year} {2013})}\BibitemShut {NoStop}%
\bibitem [{\citenamefont {Betan}\ and\ \citenamefont
  {Nazarewicz}(2012)}]{Betan2012}%
  \BibitemOpen
  \bibfield  {author} {\bibinfo {author} {\bibfnamefont {R.~I.}\ \bibnamefont
  {Betan}}\ and\ \bibinfo {author} {\bibfnamefont {W.}~\bibnamefont
  {Nazarewicz}},\ }\href {\doibase 10.1103/PhysRevC.86.034338} {\bibfield
  {journal} {\bibinfo  {journal} {Phys. Rev. C}\ }\textbf {\bibinfo {volume}
  {86}},\ \bibinfo {pages} {034338} (\bibinfo {year} {2012})}\BibitemShut
  {NoStop}%
\bibitem [{\citenamefont {Farooq-Smith}\ \emph {et~al.}(2016)\citenamefont
  {Farooq-Smith}, \citenamefont {Cocolios}, \citenamefont {Billowes},
  \citenamefont {Bissell}, \citenamefont {Budin\ifmmode \check{c}\else
  \v{c}\fi{}evi\ifmmode~\acute{c}\else \'{c}\fi{}}, \citenamefont
  {Day~Goodacre}, \citenamefont {de~Groote}, \citenamefont {Fedosseev},
  \citenamefont {Flanagan}, \citenamefont {Franchoo}, \citenamefont
  {Garcia~Ruiz}, \citenamefont {Heylen}, \citenamefont {Li}, \citenamefont
  {Lynch}, \citenamefont {Marsh}, \citenamefont {Neyens}, \citenamefont
  {Rossel}, \citenamefont {Rothe}, \citenamefont {Stroke}, \citenamefont
  {Wendt}, \citenamefont {Wilkins},\ and\ \citenamefont {Yang}}]{Flaroog2016}%
  \BibitemOpen
  \bibfield  {author} {\bibinfo {author} {\bibfnamefont {G.~J.}\ \bibnamefont
  {Farooq-Smith}}, \bibinfo {author} {\bibfnamefont {T.~E.}\ \bibnamefont
  {Cocolios}}, \bibinfo {author} {\bibfnamefont {J.}~\bibnamefont {Billowes}},
  \bibinfo {author} {\bibfnamefont {M.~L.}\ \bibnamefont {Bissell}}, \bibinfo
  {author} {\bibfnamefont {I.}~\bibnamefont {Budin\ifmmode \check{c}\else
  \v{c}\fi{}evi\ifmmode~\acute{c}\else \'{c}\fi{}}}, \bibinfo {author}
  {\bibfnamefont {T.}~\bibnamefont {Day~Goodacre}}, \bibinfo {author}
  {\bibfnamefont {R.~P.}\ \bibnamefont {de~Groote}}, \bibinfo {author}
  {\bibfnamefont {V.~N.}\ \bibnamefont {Fedosseev}}, \bibinfo {author}
  {\bibfnamefont {K.~T.}\ \bibnamefont {Flanagan}}, \bibinfo {author}
  {\bibfnamefont {S.}~\bibnamefont {Franchoo}}, \bibinfo {author}
  {\bibfnamefont {R.~F.}\ \bibnamefont {Garcia~Ruiz}}, \bibinfo {author}
  {\bibfnamefont {H.}~\bibnamefont {Heylen}}, \bibinfo {author} {\bibfnamefont
  {R.}~\bibnamefont {Li}}, \bibinfo {author} {\bibfnamefont {K.~M.}\
  \bibnamefont {Lynch}}, \bibinfo {author} {\bibfnamefont {B.~A.}\ \bibnamefont
  {Marsh}}, \bibinfo {author} {\bibfnamefont {G.}~\bibnamefont {Neyens}},
  \bibinfo {author} {\bibfnamefont {R.~E.}\ \bibnamefont {Rossel}}, \bibinfo
  {author} {\bibfnamefont {S.}~\bibnamefont {Rothe}}, \bibinfo {author}
  {\bibfnamefont {H.~H.}\ \bibnamefont {Stroke}}, \bibinfo {author}
  {\bibfnamefont {K.~D.~A.}\ \bibnamefont {Wendt}}, \bibinfo {author}
  {\bibfnamefont {S.~G.}\ \bibnamefont {Wilkins}}, \ and\ \bibinfo {author}
  {\bibfnamefont {X.~F.}\ \bibnamefont {Yang}},\ }\href {\doibase
  10.1103/PhysRevC.94.054305} {\bibfield  {journal} {\bibinfo  {journal} {Phys.
  Rev. C}\ }\textbf {\bibinfo {volume} {94}},\ \bibinfo {pages} {054305}
  (\bibinfo {year} {2016})}\BibitemShut {NoStop}%
\bibitem [{\citenamefont {Andreyev}\ \emph {et~al.}(1992)\citenamefont
  {Andreyev}, \citenamefont {Bogdanov}, \citenamefont {Chepigin}, \citenamefont
  {Kabachenko}, \citenamefont {Malyshev}, \citenamefont {Sagajdak},
  \citenamefont {Ter-Akopian},\ and\ \citenamefont {Yeremin}}]{Andreyev1992}%
  \BibitemOpen
  \bibfield  {author} {\bibinfo {author} {\bibfnamefont {A.~N.}\ \bibnamefont
  {Andreyev}}, \bibinfo {author} {\bibfnamefont {D.~D.}\ \bibnamefont
  {Bogdanov}}, \bibinfo {author} {\bibfnamefont {V.~I.}\ \bibnamefont
  {Chepigin}}, \bibinfo {author} {\bibfnamefont {A.~P.}\ \bibnamefont
  {Kabachenko}}, \bibinfo {author} {\bibfnamefont {O.~N.}\ \bibnamefont
  {Malyshev}}, \bibinfo {author} {\bibfnamefont {R.~N.}\ \bibnamefont
  {Sagajdak}}, \bibinfo {author} {\bibfnamefont {G.~M.}\ \bibnamefont
  {Ter-Akopian}}, \ and\ \bibinfo {author} {\bibfnamefont {A.~V.}\ \bibnamefont
  {Yeremin}},\ }\href {\doibase 10.1007/BF01294498} {\bibfield  {journal}
  {\bibinfo  {journal} {Z. Phys. A}\ }\textbf {\bibinfo {volume} {342}},\
  \bibinfo {pages} {123} (\bibinfo {year} {1992})}\BibitemShut {NoStop}%
\bibitem [{\citenamefont {Andreyev}\ \emph {et~al.}(1993)\citenamefont
  {Andreyev}, \citenamefont {Bogdanov}, \citenamefont {Chepigin}, \citenamefont
  {Kabachenko}, \citenamefont {Malyshev}, \citenamefont {Sagaidak},
  \citenamefont {Ter-Akopian}, \citenamefont {Veselsky},\ and\ \citenamefont
  {Yeremin}}]{Andreyev1993}%
  \BibitemOpen
  \bibfield  {author} {\bibinfo {author} {\bibfnamefont {A.~N.}\ \bibnamefont
  {Andreyev}}, \bibinfo {author} {\bibfnamefont {D.~D.}\ \bibnamefont
  {Bogdanov}}, \bibinfo {author} {\bibfnamefont {V.~I.}\ \bibnamefont
  {Chepigin}}, \bibinfo {author} {\bibfnamefont {A.~P.}\ \bibnamefont
  {Kabachenko}}, \bibinfo {author} {\bibfnamefont {O.~N.}\ \bibnamefont
  {Malyshev}}, \bibinfo {author} {\bibfnamefont {R.~N.}\ \bibnamefont
  {Sagaidak}}, \bibinfo {author} {\bibfnamefont {G.~M.}\ \bibnamefont
  {Ter-Akopian}}, \bibinfo {author} {\bibfnamefont {M.}~\bibnamefont
  {Veselsky}}, \ and\ \bibinfo {author} {\bibfnamefont {A.~V.}\ \bibnamefont
  {Yeremin}},\ }\href {\doibase 10.1007/BF01293353} {\bibfield  {journal}
  {\bibinfo  {journal} {Z. Phys. A}\ }\textbf {\bibinfo {volume} {345}},\
  \bibinfo {pages} {247} (\bibinfo {year} {1993})}\BibitemShut {NoStop}%
\bibitem [{\citenamefont {Lepp\"anen}\ \emph {et~al.}(2007)\citenamefont
  {Lepp\"anen}, \citenamefont {Uusitalo}, \citenamefont {Leino}, \citenamefont
  {Eeckhaudt}, \citenamefont {Grahn}, \citenamefont {Greenlees}, \citenamefont
  {Jones}, \citenamefont {Julin}, \citenamefont {Juutinen}, \citenamefont
  {Kettunen}, \citenamefont {Kuusiniemi}, \citenamefont {Nieminen},
  \citenamefont {Pakarinen}, \citenamefont {Rahkila}, \citenamefont {Scholey},\
  and\ \citenamefont {Sletten}}]{Lepanen2007}%
  \BibitemOpen
  \bibfield  {author} {\bibinfo {author} {\bibfnamefont {A.~P.}\ \bibnamefont
  {Lepp\"anen}}, \bibinfo {author} {\bibfnamefont {J.}~\bibnamefont
  {Uusitalo}}, \bibinfo {author} {\bibfnamefont {M.}~\bibnamefont {Leino}},
  \bibinfo {author} {\bibfnamefont {S.}~\bibnamefont {Eeckhaudt}}, \bibinfo
  {author} {\bibfnamefont {T.}~\bibnamefont {Grahn}}, \bibinfo {author}
  {\bibfnamefont {P.~T.}\ \bibnamefont {Greenlees}}, \bibinfo {author}
  {\bibfnamefont {P.}~\bibnamefont {Jones}}, \bibinfo {author} {\bibfnamefont
  {R.}~\bibnamefont {Julin}}, \bibinfo {author} {\bibfnamefont
  {S.}~\bibnamefont {Juutinen}}, \bibinfo {author} {\bibfnamefont
  {H.}~\bibnamefont {Kettunen}}, \bibinfo {author} {\bibfnamefont
  {P.}~\bibnamefont {Kuusiniemi}}, \bibinfo {author} {\bibfnamefont
  {P.}~\bibnamefont {Nieminen}}, \bibinfo {author} {\bibfnamefont
  {J.}~\bibnamefont {Pakarinen}}, \bibinfo {author} {\bibfnamefont
  {P.}~\bibnamefont {Rahkila}}, \bibinfo {author} {\bibfnamefont
  {C.}~\bibnamefont {Scholey}}, \ and\ \bibinfo {author} {\bibfnamefont
  {G.}~\bibnamefont {Sletten}},\ }\href {\doibase 10.1103/PhysRevC.75.054307}
  {\bibfield  {journal} {\bibinfo  {journal} {Phys. Rev. C}\ }\textbf {\bibinfo
  {volume} {75}},\ \bibinfo {pages} {054307} (\bibinfo {year}
  {2007})}\BibitemShut {NoStop}%
\bibitem [{\citenamefont {Khuyagbaatar}\ \emph {et~al.}(2015)\citenamefont
  {Khuyagbaatar}, \citenamefont {Yakushev}, \citenamefont {D\"ullmann},
  \citenamefont {Ackermann}, \citenamefont {Andersson}, \citenamefont {Block},
  \citenamefont {Brand}, \citenamefont {Cox}, \citenamefont {Even},
  \citenamefont {Forsberg}, \citenamefont {Golubev}, \citenamefont {Hartmann},
  \citenamefont {Herzberg}, \citenamefont {He\ss{}berger}, \citenamefont
  {Hoffmann}, \citenamefont {H\"ubner}, \citenamefont {J\"ager}, \citenamefont
  {Jeppsson}, \citenamefont {Kindler}, \citenamefont {Kratz}, \citenamefont
  {Krier}, \citenamefont {Kurz}, \citenamefont {Lommel}, \citenamefont {Maiti},
  \citenamefont {Minami}, \citenamefont {Mistry}, \citenamefont {Mrosek},
  \citenamefont {Pysmenetska}, \citenamefont {Rudolph}, \citenamefont
  {Sarmiento}, \citenamefont {Schaffner}, \citenamefont {Sch\"adel},
  \citenamefont {Schausten}, \citenamefont {Steiner}, \citenamefont
  {De~Heidenreich}, \citenamefont {Uusitalo}, \citenamefont {Wegrzecki},
  \citenamefont {Wiehl},\ and\ \citenamefont {Yakusheva}}]{Khuyagbaatar2015}%
  \BibitemOpen
  \bibfield  {author} {\bibinfo {author} {\bibfnamefont {J.}~\bibnamefont
  {Khuyagbaatar}}, \bibinfo {author} {\bibfnamefont {A.}~\bibnamefont
  {Yakushev}}, \bibinfo {author} {\bibfnamefont {C.~E.}\ \bibnamefont
  {D\"ullmann}}, \bibinfo {author} {\bibfnamefont {D.}~\bibnamefont
  {Ackermann}}, \bibinfo {author} {\bibfnamefont {L.-L.}\ \bibnamefont
  {Andersson}}, \bibinfo {author} {\bibfnamefont {M.}~\bibnamefont {Block}},
  \bibinfo {author} {\bibfnamefont {H.}~\bibnamefont {Brand}}, \bibinfo
  {author} {\bibfnamefont {D.~M.}\ \bibnamefont {Cox}}, \bibinfo {author}
  {\bibfnamefont {J.}~\bibnamefont {Even}}, \bibinfo {author} {\bibfnamefont
  {U.}~\bibnamefont {Forsberg}}, \bibinfo {author} {\bibfnamefont
  {P.}~\bibnamefont {Golubev}}, \bibinfo {author} {\bibfnamefont
  {W.}~\bibnamefont {Hartmann}}, \bibinfo {author} {\bibfnamefont {R.-D.}\
  \bibnamefont {Herzberg}}, \bibinfo {author} {\bibfnamefont {F.~P.}\
  \bibnamefont {He\ss{}berger}}, \bibinfo {author} {\bibfnamefont
  {J.}~\bibnamefont {Hoffmann}}, \bibinfo {author} {\bibfnamefont
  {A.}~\bibnamefont {H\"ubner}}, \bibinfo {author} {\bibfnamefont
  {E.}~\bibnamefont {J\"ager}}, \bibinfo {author} {\bibfnamefont
  {J.}~\bibnamefont {Jeppsson}}, \bibinfo {author} {\bibfnamefont
  {B.}~\bibnamefont {Kindler}}, \bibinfo {author} {\bibfnamefont {J.~V.}\
  \bibnamefont {Kratz}}, \bibinfo {author} {\bibfnamefont {J.}~\bibnamefont
  {Krier}}, \bibinfo {author} {\bibfnamefont {N.}~\bibnamefont {Kurz}},
  \bibinfo {author} {\bibfnamefont {B.}~\bibnamefont {Lommel}}, \bibinfo
  {author} {\bibfnamefont {M.}~\bibnamefont {Maiti}}, \bibinfo {author}
  {\bibfnamefont {S.}~\bibnamefont {Minami}}, \bibinfo {author} {\bibfnamefont
  {A.~K.}\ \bibnamefont {Mistry}}, \bibinfo {author} {\bibfnamefont {C.~M.}\
  \bibnamefont {Mrosek}}, \bibinfo {author} {\bibfnamefont {I.}~\bibnamefont
  {Pysmenetska}}, \bibinfo {author} {\bibfnamefont {D.}~\bibnamefont
  {Rudolph}}, \bibinfo {author} {\bibfnamefont {L.~G.}\ \bibnamefont
  {Sarmiento}}, \bibinfo {author} {\bibfnamefont {H.}~\bibnamefont
  {Schaffner}}, \bibinfo {author} {\bibfnamefont {M.}~\bibnamefont
  {Sch\"adel}}, \bibinfo {author} {\bibfnamefont {B.}~\bibnamefont
  {Schausten}}, \bibinfo {author} {\bibfnamefont {J.}~\bibnamefont {Steiner}},
  \bibinfo {author} {\bibfnamefont {T.~T.}\ \bibnamefont {De~Heidenreich}},
  \bibinfo {author} {\bibfnamefont {J.}~\bibnamefont {Uusitalo}}, \bibinfo
  {author} {\bibfnamefont {M.}~\bibnamefont {Wegrzecki}}, \bibinfo {author}
  {\bibfnamefont {N.}~\bibnamefont {Wiehl}}, \ and\ \bibinfo {author}
  {\bibfnamefont {V.}~\bibnamefont {Yakusheva}},\ }\href {\doibase
  10.1103/PhysRevLett.115.242502} {\bibfield  {journal} {\bibinfo  {journal}
  {Phys. Rev. Lett.}\ }\textbf {\bibinfo {volume} {115}},\ \bibinfo {pages}
  {242502} (\bibinfo {year} {2015})}\BibitemShut {NoStop}%
\bibitem [{\citenamefont {Yang}\ \emph {et~al.}(2015)\citenamefont {Yang},
  \citenamefont {Zhang}, \citenamefont {Wang}, \citenamefont {Gan},
  \citenamefont {Ma}, \citenamefont {Yu}, \citenamefont {Jiang}, \citenamefont
  {Tian}, \citenamefont {Ding}, \citenamefont {Guo}, \citenamefont {Wang},
  \citenamefont {Huang}, \citenamefont {Sun}, \citenamefont {Wang},
  \citenamefont {Zhou}, \citenamefont {Ren}, \citenamefont {Zhou},
  \citenamefont {Xu},\ and\ \citenamefont {Xiao}}]{Yang2015}%
  \BibitemOpen
  \bibfield  {author} {\bibinfo {author} {\bibfnamefont {H.~B.}\ \bibnamefont
  {Yang}}, \bibinfo {author} {\bibfnamefont {Z.~Y.}\ \bibnamefont {Zhang}},
  \bibinfo {author} {\bibfnamefont {J.~G.}\ \bibnamefont {Wang}}, \bibinfo
  {author} {\bibfnamefont {Z.~G.}\ \bibnamefont {Gan}}, \bibinfo {author}
  {\bibfnamefont {L.}~\bibnamefont {Ma}}, \bibinfo {author} {\bibfnamefont
  {L.}~\bibnamefont {Yu}}, \bibinfo {author} {\bibfnamefont {J.}~\bibnamefont
  {Jiang}}, \bibinfo {author} {\bibfnamefont {Y.~L.}\ \bibnamefont {Tian}},
  \bibinfo {author} {\bibfnamefont {B.}~\bibnamefont {Ding}}, \bibinfo {author}
  {\bibfnamefont {S.}~\bibnamefont {Guo}}, \bibinfo {author} {\bibfnamefont
  {Y.~S.}\ \bibnamefont {Wang}}, \bibinfo {author} {\bibfnamefont {T.~H.}\
  \bibnamefont {Huang}}, \bibinfo {author} {\bibfnamefont {M.~D.}\ \bibnamefont
  {Sun}}, \bibinfo {author} {\bibfnamefont {K.~L.}\ \bibnamefont {Wang}},
  \bibinfo {author} {\bibfnamefont {S.~G.}\ \bibnamefont {Zhou}}, \bibinfo
  {author} {\bibfnamefont {Z.~Z.}\ \bibnamefont {Ren}}, \bibinfo {author}
  {\bibfnamefont {X.~H.}\ \bibnamefont {Zhou}}, \bibinfo {author}
  {\bibfnamefont {H.~S.}\ \bibnamefont {Xu}}, \ and\ \bibinfo {author}
  {\bibfnamefont {G.~Q.}\ \bibnamefont {Xiao}},\ }\href {\doibase
  10.1140/epja/i2015-15088-9} {\bibfield  {journal} {\bibinfo  {journal} {Eur.
  Phys. J. A}\ }\textbf {\bibinfo {volume} {51}},\ \bibinfo {pages} {88}
  (\bibinfo {year} {2015})}\BibitemShut {NoStop}%
\bibitem [{\citenamefont {Sun}\ \emph {et~al.}(2017)\citenamefont {Sun},
  \citenamefont {Liu}, \citenamefont {Huang}, \citenamefont {Zhang},
  \citenamefont {Wang}, \citenamefont {Liu}, \citenamefont {Ding},
  \citenamefont {Gan}, \citenamefont {Ma}, \citenamefont {Yang}, \citenamefont
  {Zhang}, \citenamefont {Yu}, \citenamefont {Jiang}, \citenamefont {Wang},
  \citenamefont {Wang}, \citenamefont {Liu}, \citenamefont {Li}, \citenamefont
  {Li}, \citenamefont {Wang}, \citenamefont {Lu}, \citenamefont {Lin},
  \citenamefont {Sun}, \citenamefont {Ma}, \citenamefont {Yuan}, \citenamefont
  {Zuo}, \citenamefont {Xu}, \citenamefont {Zhou}, \citenamefont {Xiao},
  \citenamefont {Qi},\ and\ \citenamefont {Zhang}}]{SUN2017303}%
  \BibitemOpen
  \bibfield  {author} {\bibinfo {author} {\bibfnamefont {M.}~\bibnamefont
  {Sun}}, \bibinfo {author} {\bibfnamefont {Z.}~\bibnamefont {Liu}}, \bibinfo
  {author} {\bibfnamefont {T.}~\bibnamefont {Huang}}, \bibinfo {author}
  {\bibfnamefont {W.}~\bibnamefont {Zhang}}, \bibinfo {author} {\bibfnamefont
  {J.}~\bibnamefont {Wang}}, \bibinfo {author} {\bibfnamefont {X.}~\bibnamefont
  {Liu}}, \bibinfo {author} {\bibfnamefont {B.}~\bibnamefont {Ding}}, \bibinfo
  {author} {\bibfnamefont {Z.}~\bibnamefont {Gan}}, \bibinfo {author}
  {\bibfnamefont {L.}~\bibnamefont {Ma}}, \bibinfo {author} {\bibfnamefont
  {H.}~\bibnamefont {Yang}}, \bibinfo {author} {\bibfnamefont {Z.}~\bibnamefont
  {Zhang}}, \bibinfo {author} {\bibfnamefont {L.}~\bibnamefont {Yu}}, \bibinfo
  {author} {\bibfnamefont {J.}~\bibnamefont {Jiang}}, \bibinfo {author}
  {\bibfnamefont {K.}~\bibnamefont {Wang}}, \bibinfo {author} {\bibfnamefont
  {Y.}~\bibnamefont {Wang}}, \bibinfo {author} {\bibfnamefont {M.}~\bibnamefont
  {Liu}}, \bibinfo {author} {\bibfnamefont {Z.}~\bibnamefont {Li}}, \bibinfo
  {author} {\bibfnamefont {J.}~\bibnamefont {Li}}, \bibinfo {author}
  {\bibfnamefont {X.}~\bibnamefont {Wang}}, \bibinfo {author} {\bibfnamefont
  {H.}~\bibnamefont {Lu}}, \bibinfo {author} {\bibfnamefont {C.}~\bibnamefont
  {Lin}}, \bibinfo {author} {\bibfnamefont {L.}~\bibnamefont {Sun}}, \bibinfo
  {author} {\bibfnamefont {N.}~\bibnamefont {Ma}}, \bibinfo {author}
  {\bibfnamefont {C.}~\bibnamefont {Yuan}}, \bibinfo {author} {\bibfnamefont
  {W.}~\bibnamefont {Zuo}}, \bibinfo {author} {\bibfnamefont {H.}~\bibnamefont
  {Xu}}, \bibinfo {author} {\bibfnamefont {X.}~\bibnamefont {Zhou}}, \bibinfo
  {author} {\bibfnamefont {G.}~\bibnamefont {Xiao}}, \bibinfo {author}
  {\bibfnamefont {C.}~\bibnamefont {Qi}}, \ and\ \bibinfo {author}
  {\bibfnamefont {F.}~\bibnamefont {Zhang}},\ }\href {\doibase
  https://doi.org/10.1016/j.physletb.2017.03.074} {\bibfield  {journal}
  {\bibinfo  {journal} {Phys. Lett. B}\ }\textbf {\bibinfo {volume} {771}},\
  \bibinfo {pages} {303 } (\bibinfo {year} {2017})}\BibitemShut {NoStop}%
\bibitem [{\citenamefont {Lunney}\ \emph {et~al.}(2003)\citenamefont {Lunney},
  \citenamefont {Pearson},\ and\ \citenamefont {Thibault}}]{Lunney2003}%
  \BibitemOpen
  \bibfield  {author} {\bibinfo {author} {\bibfnamefont {D.}~\bibnamefont
  {Lunney}}, \bibinfo {author} {\bibfnamefont {J.~M.}\ \bibnamefont {Pearson}},
  \ and\ \bibinfo {author} {\bibfnamefont {C.}~\bibnamefont {Thibault}},\
  }\href {\doibase 10.1103/RevModPhys.75.1021} {\bibfield  {journal} {\bibinfo
  {journal} {Rev. Mod. Phys.}\ }\textbf {\bibinfo {volume} {75}},\ \bibinfo
  {pages} {1021} (\bibinfo {year} {2003})}\BibitemShut {NoStop}%
\bibitem [{\citenamefont {Huang}\ and\ \citenamefont {Audi}(2017)}]{SF2017}%
  \BibitemOpen
  \bibfield  {author} {\bibinfo {author} {\bibfnamefont {W.}~\bibnamefont
  {Huang}}\ and\ \bibinfo {author} {\bibfnamefont {G.}~\bibnamefont {Audi}},\
  }\href@noop {} {\bibfield  {journal} {\bibinfo  {journal}
  {arXiv:1702.01639v1}\ } (\bibinfo {year} {2017})}\BibitemShut {NoStop}%
\bibitem [{\citenamefont {Wang}\ \emph {et~al.}(2017)\citenamefont {Wang},
  \citenamefont {Audi}, \citenamefont {Kondev}, \citenamefont {Huang},
  \citenamefont {Naimi},\ and\ \citenamefont {Xu}}]{AME2016}%
  \BibitemOpen
  \bibfield  {author} {\bibinfo {author} {\bibfnamefont {M.}~\bibnamefont
  {Wang}}, \bibinfo {author} {\bibfnamefont {G.}~\bibnamefont {Audi}}, \bibinfo
  {author} {\bibfnamefont {F.}~\bibnamefont {Kondev}}, \bibinfo {author}
  {\bibfnamefont {W.}~\bibnamefont {Huang}}, \bibinfo {author} {\bibfnamefont
  {S.}~\bibnamefont {Naimi}}, \ and\ \bibinfo {author} {\bibfnamefont
  {X.}~\bibnamefont {Xu}},\ }\href
  {http://stacks.iop.org/1674-1137/41/i=3/a=030003} {\bibfield  {journal}
  {\bibinfo  {journal} {Chinese Physics C}\ }\textbf {\bibinfo {volume} {41}},\
  \bibinfo {pages} {030003} (\bibinfo {year} {2017})}\BibitemShut {NoStop}%
\bibitem [{\citenamefont {Alkhazov}\ \emph {et~al.}(1983)\citenamefont
  {Alkhazov}, \citenamefont {Mezilev}, \citenamefont {Novikov}, \citenamefont
  {Ganbaatar}, \citenamefont {Gromov}, \citenamefont {Kalinnikov},
  \citenamefont {Potempa}, \citenamefont {Sieniawski},\ and\ \citenamefont
  {Tarkanyi}}]{Alkhazov1983}%
  \BibitemOpen
  \bibfield  {author} {\bibinfo {author} {\bibfnamefont {G.~D.}\ \bibnamefont
  {Alkhazov}}, \bibinfo {author} {\bibfnamefont {K.~A.}\ \bibnamefont
  {Mezilev}}, \bibinfo {author} {\bibfnamefont {Y.~N.}\ \bibnamefont
  {Novikov}}, \bibinfo {author} {\bibfnamefont {N.}~\bibnamefont {Ganbaatar}},
  \bibinfo {author} {\bibfnamefont {K.~Y.}\ \bibnamefont {Gromov}}, \bibinfo
  {author} {\bibfnamefont {V.~G.}\ \bibnamefont {Kalinnikov}}, \bibinfo
  {author} {\bibfnamefont {A.}~\bibnamefont {Potempa}}, \bibinfo {author}
  {\bibfnamefont {E.}~\bibnamefont {Sieniawski}}, \ and\ \bibinfo {author}
  {\bibfnamefont {F.}~\bibnamefont {Tarkanyi}},\ }\href {\doibase
  10.1007/BF01415230} {\bibfield  {journal} {\bibinfo  {journal} {Zeitschrift
  f{\"u}r Physik A Atoms and Nuclei}\ }\textbf {\bibinfo {volume} {310}},\
  \bibinfo {pages} {247} (\bibinfo {year} {1983})}\BibitemShut {NoStop}%
\bibitem [{\citenamefont {Beck}\ \emph {et~al.}(2000)\citenamefont {Beck},
  \citenamefont {Ames}, \citenamefont {Audi}, \citenamefont {Bollen},
  \citenamefont {Herfurth}, \citenamefont {Kluge}, \citenamefont {Kohl},
  \citenamefont {K{\"o}nig}, \citenamefont {Lunney}, \citenamefont {Martel},
  \citenamefont {Moore}, \citenamefont {Raimbault-Hartmann}, \citenamefont
  {Schark}, \citenamefont {Schwarz}, \citenamefont {de~Saint~Simon},\ and\
  \citenamefont {Szerypo}}]{Beck2000a}%
  \BibitemOpen
  \bibfield  {author} {\bibinfo {author} {\bibfnamefont {D.}~\bibnamefont
  {Beck}}, \bibinfo {author} {\bibfnamefont {F.}~\bibnamefont {Ames}}, \bibinfo
  {author} {\bibfnamefont {G.}~\bibnamefont {Audi}}, \bibinfo {author}
  {\bibfnamefont {G.}~\bibnamefont {Bollen}}, \bibinfo {author} {\bibfnamefont
  {F.}~\bibnamefont {Herfurth}}, \bibinfo {author} {\bibfnamefont {H.~J.}\
  \bibnamefont {Kluge}}, \bibinfo {author} {\bibfnamefont {A.}~\bibnamefont
  {Kohl}}, \bibinfo {author} {\bibfnamefont {M.}~\bibnamefont {K{\"o}nig}},
  \bibinfo {author} {\bibfnamefont {D.}~\bibnamefont {Lunney}}, \bibinfo
  {author} {\bibfnamefont {I.}~\bibnamefont {Martel}}, \bibinfo {author}
  {\bibfnamefont {R.~B.}\ \bibnamefont {Moore}}, \bibinfo {author}
  {\bibfnamefont {H.}~\bibnamefont {Raimbault-Hartmann}}, \bibinfo {author}
  {\bibfnamefont {E.}~\bibnamefont {Schark}}, \bibinfo {author} {\bibfnamefont
  {S.}~\bibnamefont {Schwarz}}, \bibinfo {author} {\bibfnamefont
  {M.}~\bibnamefont {de~Saint~Simon}}, \ and\ \bibinfo {author} {\bibfnamefont
  {J.}~\bibnamefont {Szerypo}},\ }\href {\doibase 10.1007/s100500070085}
  {\bibfield  {journal} {\bibinfo  {journal} {Eur. Phys. J. A}\ }\textbf
  {\bibinfo {volume} {8}},\ \bibinfo {pages} {307} (\bibinfo {year}
  {2000})}\BibitemShut {NoStop}%
\bibitem [{\citenamefont {Stanja}\ \emph {et~al.}(2013)\citenamefont {Stanja},
  \citenamefont {Borgmann}, \citenamefont {Agramunt}, \citenamefont {Algora},
  \citenamefont {Beck}, \citenamefont {Blaum}, \citenamefont {B\"ohm},
  \citenamefont {Breitenfeldt}, \citenamefont {Cocolios}, \citenamefont
  {Fraile}, \citenamefont {Herfurth}, \citenamefont {Herlert}, \citenamefont
  {Kowalska}, \citenamefont {Kreim}, \citenamefont {Lunney}, \citenamefont
  {Manea}, \citenamefont {Minaya~Ramirez}, \citenamefont {Naimi}, \citenamefont
  {Neidherr}, \citenamefont {Rosenbusch}, \citenamefont {Schweikhard},
  \citenamefont {Simpson}, \citenamefont {Wienholtz}, \citenamefont {Wolf},\
  and\ \citenamefont {Zuber}}]{Stanja2013}%
  \BibitemOpen
  \bibfield  {author} {\bibinfo {author} {\bibfnamefont {J.}~\bibnamefont
  {Stanja}}, \bibinfo {author} {\bibfnamefont {C.}~\bibnamefont {Borgmann}},
  \bibinfo {author} {\bibfnamefont {J.}~\bibnamefont {Agramunt}}, \bibinfo
  {author} {\bibfnamefont {A.}~\bibnamefont {Algora}}, \bibinfo {author}
  {\bibfnamefont {D.}~\bibnamefont {Beck}}, \bibinfo {author} {\bibfnamefont
  {K.}~\bibnamefont {Blaum}}, \bibinfo {author} {\bibfnamefont
  {C.}~\bibnamefont {B\"ohm}}, \bibinfo {author} {\bibfnamefont
  {M.}~\bibnamefont {Breitenfeldt}}, \bibinfo {author} {\bibfnamefont {T.~E.}\
  \bibnamefont {Cocolios}}, \bibinfo {author} {\bibfnamefont {L.~M.}\
  \bibnamefont {Fraile}}, \bibinfo {author} {\bibfnamefont {F.}~\bibnamefont
  {Herfurth}}, \bibinfo {author} {\bibfnamefont {A.}~\bibnamefont {Herlert}},
  \bibinfo {author} {\bibfnamefont {M.}~\bibnamefont {Kowalska}}, \bibinfo
  {author} {\bibfnamefont {S.}~\bibnamefont {Kreim}}, \bibinfo {author}
  {\bibfnamefont {D.}~\bibnamefont {Lunney}}, \bibinfo {author} {\bibfnamefont
  {V.}~\bibnamefont {Manea}}, \bibinfo {author} {\bibfnamefont
  {E.}~\bibnamefont {Minaya~Ramirez}}, \bibinfo {author} {\bibfnamefont
  {S.}~\bibnamefont {Naimi}}, \bibinfo {author} {\bibfnamefont
  {D.}~\bibnamefont {Neidherr}}, \bibinfo {author} {\bibfnamefont
  {M.}~\bibnamefont {Rosenbusch}}, \bibinfo {author} {\bibfnamefont
  {L.}~\bibnamefont {Schweikhard}}, \bibinfo {author} {\bibfnamefont
  {G.}~\bibnamefont {Simpson}}, \bibinfo {author} {\bibfnamefont
  {F.}~\bibnamefont {Wienholtz}}, \bibinfo {author} {\bibfnamefont {R.~N.}\
  \bibnamefont {Wolf}}, \ and\ \bibinfo {author} {\bibfnamefont
  {K.}~\bibnamefont {Zuber}},\ }\href {\doibase 10.1103/PhysRevC.88.054304}
  {\bibfield  {journal} {\bibinfo  {journal} {Phys. Rev. C}\ }\textbf {\bibinfo
  {volume} {88}},\ \bibinfo {pages} {054304} (\bibinfo {year}
  {2013})}\BibitemShut {NoStop}%
\bibitem [{\citenamefont {Althubiti}\ \emph {et~al.}(2017)\citenamefont
  {Althubiti}, \citenamefont {Atanasov}, \citenamefont {Blaum}, \citenamefont
  {Cocolios}, \citenamefont {Day~Goodacre}, \citenamefont {Farooq-Smith},
  \citenamefont {Fedorov}, \citenamefont {Fedosseev}, \citenamefont {George},
  \citenamefont {Herfurth}, \citenamefont {Heyde}, \citenamefont {Kreim},
  \citenamefont {Lunney}, \citenamefont {Lynch}, \citenamefont {Manea},
  \citenamefont {Marsh}, \citenamefont {Neidherr}, \citenamefont {Rosenbusch},
  \citenamefont {Rossel}, \citenamefont {Rothe}, \citenamefont {Schweikhard},
  \citenamefont {Seliverstov}, \citenamefont {Welker}, \citenamefont
  {Wienholtz}, \citenamefont {Wolf},\ and\ \citenamefont {Zuber}}]{Numa2017}%
  \BibitemOpen
  \bibfield  {author} {\bibinfo {author} {\bibfnamefont {N.~A.}\ \bibnamefont
  {Althubiti}}, \bibinfo {author} {\bibfnamefont {D.}~\bibnamefont {Atanasov}},
  \bibinfo {author} {\bibfnamefont {K.}~\bibnamefont {Blaum}}, \bibinfo
  {author} {\bibfnamefont {T.~E.}\ \bibnamefont {Cocolios}}, \bibinfo {author}
  {\bibfnamefont {T.}~\bibnamefont {Day~Goodacre}}, \bibinfo {author}
  {\bibfnamefont {G.~J.}\ \bibnamefont {Farooq-Smith}}, \bibinfo {author}
  {\bibfnamefont {D.~V.}\ \bibnamefont {Fedorov}}, \bibinfo {author}
  {\bibfnamefont {V.~N.}\ \bibnamefont {Fedosseev}}, \bibinfo {author}
  {\bibfnamefont {S.}~\bibnamefont {George}}, \bibinfo {author} {\bibfnamefont
  {F.}~\bibnamefont {Herfurth}}, \bibinfo {author} {\bibfnamefont
  {K.}~\bibnamefont {Heyde}}, \bibinfo {author} {\bibfnamefont
  {S.}~\bibnamefont {Kreim}}, \bibinfo {author} {\bibfnamefont
  {D.}~\bibnamefont {Lunney}}, \bibinfo {author} {\bibfnamefont {K.~M.}\
  \bibnamefont {Lynch}}, \bibinfo {author} {\bibfnamefont {V.}~\bibnamefont
  {Manea}}, \bibinfo {author} {\bibfnamefont {B.~A.}\ \bibnamefont {Marsh}},
  \bibinfo {author} {\bibfnamefont {D.}~\bibnamefont {Neidherr}}, \bibinfo
  {author} {\bibfnamefont {M.}~\bibnamefont {Rosenbusch}}, \bibinfo {author}
  {\bibfnamefont {R.~E.}\ \bibnamefont {Rossel}}, \bibinfo {author}
  {\bibfnamefont {S.}~\bibnamefont {Rothe}}, \bibinfo {author} {\bibfnamefont
  {L.}~\bibnamefont {Schweikhard}}, \bibinfo {author} {\bibfnamefont {M.~D.}\
  \bibnamefont {Seliverstov}}, \bibinfo {author} {\bibfnamefont
  {A.}~\bibnamefont {Welker}}, \bibinfo {author} {\bibfnamefont
  {F.}~\bibnamefont {Wienholtz}}, \bibinfo {author} {\bibfnamefont {R.~N.}\
  \bibnamefont {Wolf}}, \ and\ \bibinfo {author} {\bibfnamefont
  {K.}~\bibnamefont {Zuber}} (\bibinfo {collaboration} {ISOLTRAP
  Collaboration}),\ }\href {\doibase 10.1103/PhysRevC.96.044325} {\bibfield
  {journal} {\bibinfo  {journal} {Phys. Rev. C}\ }\textbf {\bibinfo {volume}
  {96}},\ \bibinfo {pages} {044325} (\bibinfo {year} {2017})}\BibitemShut
  {NoStop}%
\bibitem [{\citenamefont {Wollnik}\ and\ \citenamefont
  {Przewloka}(1990)}]{WOLLNIK1990267}%
  \BibitemOpen
  \bibfield  {author} {\bibinfo {author} {\bibfnamefont {H.}~\bibnamefont
  {Wollnik}}\ and\ \bibinfo {author} {\bibfnamefont {M.}~\bibnamefont
  {Przewloka}},\ }\href {\doibase https://doi.org/10.1016/0168-1176(90)85127-N}
  {\bibfield  {journal} {\bibinfo  {journal} {Int. J. Mass Spectrom. Ion
  Proc.}\ }\textbf {\bibinfo {volume} {96}},\ \bibinfo {pages} {267 } (\bibinfo
  {year} {1990})}\BibitemShut {NoStop}%
\bibitem [{\citenamefont {Schury}\ \emph
  {et~al.}(2014{\natexlab{a}})\citenamefont {Schury}, \citenamefont {Ito},
  \citenamefont {Wada},\ and\ \citenamefont {Wollnik}}]{SCHURY201419}%
  \BibitemOpen
  \bibfield  {author} {\bibinfo {author} {\bibfnamefont {P.}~\bibnamefont
  {Schury}}, \bibinfo {author} {\bibfnamefont {Y.}~\bibnamefont {Ito}},
  \bibinfo {author} {\bibfnamefont {M.}~\bibnamefont {Wada}}, \ and\ \bibinfo
  {author} {\bibfnamefont {H.}~\bibnamefont {Wollnik}},\ }\href {\doibase
  https://doi.org/10.1016/j.ijms.2013.11.005} {\bibfield  {journal} {\bibinfo
  {journal} {Int. J. Mass Spectrom.}\ }\textbf {\bibinfo {volume} {359}},\
  \bibinfo {pages} {19 } (\bibinfo {year} {2014}{\natexlab{a}})}\BibitemShut
  {NoStop}%
\bibitem [{\citenamefont {Schury}\ \emph
  {et~al.}(2014{\natexlab{b}})\citenamefont {Schury}, \citenamefont {Wada},
  \citenamefont {Ito}, \citenamefont {Arai}, \citenamefont {Naimi},
  \citenamefont {Sonoda}, \citenamefont {Wollnik}, \citenamefont {Shchepunov},
  \citenamefont {Smorra},\ and\ \citenamefont {Yuan}}]{SCHURY201439}%
  \BibitemOpen
  \bibfield  {author} {\bibinfo {author} {\bibfnamefont {P.}~\bibnamefont
  {Schury}}, \bibinfo {author} {\bibfnamefont {M.}~\bibnamefont {Wada}},
  \bibinfo {author} {\bibfnamefont {Y.}~\bibnamefont {Ito}}, \bibinfo {author}
  {\bibfnamefont {F.}~\bibnamefont {Arai}}, \bibinfo {author} {\bibfnamefont
  {S.}~\bibnamefont {Naimi}}, \bibinfo {author} {\bibfnamefont
  {T.}~\bibnamefont {Sonoda}}, \bibinfo {author} {\bibfnamefont
  {H.}~\bibnamefont {Wollnik}}, \bibinfo {author} {\bibfnamefont
  {V.}~\bibnamefont {Shchepunov}}, \bibinfo {author} {\bibfnamefont
  {C.}~\bibnamefont {Smorra}}, \ and\ \bibinfo {author} {\bibfnamefont
  {C.}~\bibnamefont {Yuan}},\ }\href {\doibase
  https://doi.org/10.1016/j.nimb.2014.05.016} {\bibfield  {journal} {\bibinfo
  {journal} {Nucl. Instr. Meth. B}\ }\textbf {\bibinfo {volume} {335}},\
  \bibinfo {pages} {39 } (\bibinfo {year} {2014}{\natexlab{b}})}\BibitemShut
  {NoStop}%
\bibitem [{\citenamefont {Schury}\ \emph {et~al.}(2016)\citenamefont {Schury},
  \citenamefont {Wada}, \citenamefont {Ito}, \citenamefont {Arai},
  \citenamefont {Kaji}, \citenamefont {Kimura}, \citenamefont {Morimoto},
  \citenamefont {Haba}, \citenamefont {Jeong}, \citenamefont {Koura},
  \citenamefont {Miyatake}, \citenamefont {Morita}, \citenamefont {Reponen},
  \citenamefont {Ozawa}, \citenamefont {Sonoda}, \citenamefont {Takamine},\
  and\ \citenamefont {Wollnik}}]{SCHURY2016425}%
  \BibitemOpen
  \bibfield  {author} {\bibinfo {author} {\bibfnamefont {P.}~\bibnamefont
  {Schury}}, \bibinfo {author} {\bibfnamefont {M.}~\bibnamefont {Wada}},
  \bibinfo {author} {\bibfnamefont {Y.}~\bibnamefont {Ito}}, \bibinfo {author}
  {\bibfnamefont {F.}~\bibnamefont {Arai}}, \bibinfo {author} {\bibfnamefont
  {D.}~\bibnamefont {Kaji}}, \bibinfo {author} {\bibfnamefont {S.}~\bibnamefont
  {Kimura}}, \bibinfo {author} {\bibfnamefont {K.}~\bibnamefont {Morimoto}},
  \bibinfo {author} {\bibfnamefont {H.}~\bibnamefont {Haba}}, \bibinfo {author}
  {\bibfnamefont {S.}~\bibnamefont {Jeong}}, \bibinfo {author} {\bibfnamefont
  {H.}~\bibnamefont {Koura}}, \bibinfo {author} {\bibfnamefont
  {H.}~\bibnamefont {Miyatake}}, \bibinfo {author} {\bibfnamefont
  {K.}~\bibnamefont {Morita}}, \bibinfo {author} {\bibfnamefont
  {M.}~\bibnamefont {Reponen}}, \bibinfo {author} {\bibfnamefont
  {A.}~\bibnamefont {Ozawa}}, \bibinfo {author} {\bibfnamefont
  {T.}~\bibnamefont {Sonoda}}, \bibinfo {author} {\bibfnamefont
  {A.}~\bibnamefont {Takamine}}, \ and\ \bibinfo {author} {\bibfnamefont
  {H.}~\bibnamefont {Wollnik}},\ }\href {\doibase
  https://doi.org/10.1016/j.nimb.2016.02.061} {\bibfield  {journal} {\bibinfo
  {journal} {Nucl. Instr. Meth. B}\ }\textbf {\bibinfo {volume} {376}},\
  \bibinfo {pages} {425 } (\bibinfo {year} {2016})}\BibitemShut {NoStop}%
\bibitem [{\citenamefont {Schury}\ \emph
  {et~al.}(2017{\natexlab{a}})\citenamefont {Schury}, \citenamefont {Wada},
  \citenamefont {Ito}, \citenamefont {Kaji}, \citenamefont {Arai},
  \citenamefont {MacCormick}, \citenamefont {Murray}, \citenamefont {Haba},
  \citenamefont {Jeong}, \citenamefont {Kimura}, \citenamefont {Koura},
  \citenamefont {Miyatake}, \citenamefont {Morimoto}, \citenamefont {Morita},
  \citenamefont {Ozawa}, \citenamefont {Rosenbusch}, \citenamefont {Reponen},
  \citenamefont {S\"oderstr\"om}, \citenamefont {Takamine}, \citenamefont
  {Tanaka},\ and\ \citenamefont {Wollnik}}]{SCHURY2017a}%
  \BibitemOpen
  \bibfield  {author} {\bibinfo {author} {\bibfnamefont {P.}~\bibnamefont
  {Schury}}, \bibinfo {author} {\bibfnamefont {M.}~\bibnamefont {Wada}},
  \bibinfo {author} {\bibfnamefont {Y.}~\bibnamefont {Ito}}, \bibinfo {author}
  {\bibfnamefont {D.}~\bibnamefont {Kaji}}, \bibinfo {author} {\bibfnamefont
  {F.}~\bibnamefont {Arai}}, \bibinfo {author} {\bibfnamefont {M.}~\bibnamefont
  {MacCormick}}, \bibinfo {author} {\bibfnamefont {I.}~\bibnamefont {Murray}},
  \bibinfo {author} {\bibfnamefont {H.}~\bibnamefont {Haba}}, \bibinfo {author}
  {\bibfnamefont {S.}~\bibnamefont {Jeong}}, \bibinfo {author} {\bibfnamefont
  {S.}~\bibnamefont {Kimura}}, \bibinfo {author} {\bibfnamefont
  {H.}~\bibnamefont {Koura}}, \bibinfo {author} {\bibfnamefont
  {H.}~\bibnamefont {Miyatake}}, \bibinfo {author} {\bibfnamefont
  {K.}~\bibnamefont {Morimoto}}, \bibinfo {author} {\bibfnamefont
  {K.}~\bibnamefont {Morita}}, \bibinfo {author} {\bibfnamefont
  {A.}~\bibnamefont {Ozawa}}, \bibinfo {author} {\bibfnamefont
  {M.}~\bibnamefont {Rosenbusch}}, \bibinfo {author} {\bibfnamefont
  {M.}~\bibnamefont {Reponen}}, \bibinfo {author} {\bibfnamefont {P.-A.}\
  \bibnamefont {S\"oderstr\"om}}, \bibinfo {author} {\bibfnamefont
  {A.}~\bibnamefont {Takamine}}, \bibinfo {author} {\bibfnamefont
  {T.}~\bibnamefont {Tanaka}}, \ and\ \bibinfo {author} {\bibfnamefont
  {H.}~\bibnamefont {Wollnik}},\ }\href {\doibase 10.1103/PhysRevC.95.011305}
  {\bibfield  {journal} {\bibinfo  {journal} {Phys. Rev. C}\ }\textbf {\bibinfo
  {volume} {95}},\ \bibinfo {pages} {011305} (\bibinfo {year}
  {2017}{\natexlab{a}})}\BibitemShut {NoStop}%
\bibitem [{\citenamefont {Ito}\ \emph {et~al.}(2018)\citenamefont {Ito},
  \citenamefont {Schury}, \citenamefont {Wada}, \citenamefont {Arai},
  \citenamefont {Haba}, \citenamefont {Hirayama}, \citenamefont {Ishizawa},
  \citenamefont {Kaji}, \citenamefont {Kimura}, \citenamefont {Koura},
  \citenamefont {MacCormick}, \citenamefont {Miyatake}, \citenamefont {Moon},
  \citenamefont {Morimoto}, \citenamefont {Morita}, \citenamefont {Mukai},
  \citenamefont {Murray}, \citenamefont {Niwase}, \citenamefont {Okada},
  \citenamefont {Ozawa}, \citenamefont {Rosenbusch}, \citenamefont {Takamine},
  \citenamefont {Tanaka}, \citenamefont {Watanabe}, \citenamefont {Wollnik},\
  and\ \citenamefont {Yamaki}}]{Ito2017}%
  \BibitemOpen
  \bibfield  {author} {\bibinfo {author} {\bibfnamefont {Y.}~\bibnamefont
  {Ito}}, \bibinfo {author} {\bibfnamefont {P.}~\bibnamefont {Schury}},
  \bibinfo {author} {\bibfnamefont {M.}~\bibnamefont {Wada}}, \bibinfo {author}
  {\bibfnamefont {F.}~\bibnamefont {Arai}}, \bibinfo {author} {\bibfnamefont
  {H.}~\bibnamefont {Haba}}, \bibinfo {author} {\bibfnamefont {Y.}~\bibnamefont
  {Hirayama}}, \bibinfo {author} {\bibfnamefont {S.}~\bibnamefont {Ishizawa}},
  \bibinfo {author} {\bibfnamefont {D.}~\bibnamefont {Kaji}}, \bibinfo {author}
  {\bibfnamefont {S.}~\bibnamefont {Kimura}}, \bibinfo {author} {\bibfnamefont
  {H.}~\bibnamefont {Koura}}, \bibinfo {author} {\bibfnamefont
  {M.}~\bibnamefont {MacCormick}}, \bibinfo {author} {\bibfnamefont
  {H.}~\bibnamefont {Miyatake}}, \bibinfo {author} {\bibfnamefont {J.~Y.}\
  \bibnamefont {Moon}}, \bibinfo {author} {\bibfnamefont {K.}~\bibnamefont
  {Morimoto}}, \bibinfo {author} {\bibfnamefont {K.}~\bibnamefont {Morita}},
  \bibinfo {author} {\bibfnamefont {M.}~\bibnamefont {Mukai}}, \bibinfo
  {author} {\bibfnamefont {I.}~\bibnamefont {Murray}}, \bibinfo {author}
  {\bibfnamefont {T.}~\bibnamefont {Niwase}}, \bibinfo {author} {\bibfnamefont
  {K.}~\bibnamefont {Okada}}, \bibinfo {author} {\bibfnamefont
  {A.}~\bibnamefont {Ozawa}}, \bibinfo {author} {\bibfnamefont
  {M.}~\bibnamefont {Rosenbusch}}, \bibinfo {author} {\bibfnamefont
  {A.}~\bibnamefont {Takamine}}, \bibinfo {author} {\bibfnamefont
  {T.}~\bibnamefont {Tanaka}}, \bibinfo {author} {\bibfnamefont {Y.~X.}\
  \bibnamefont {Watanabe}}, \bibinfo {author} {\bibfnamefont {H.}~\bibnamefont
  {Wollnik}}, \ and\ \bibinfo {author} {\bibfnamefont {S.}~\bibnamefont
  {Yamaki}},\ }\href {\doibase 10.1103/PhysRevLett.120.152501} {\bibfield
  {journal} {\bibinfo  {journal} {Phys. Rev. Lett.}\ }\textbf {\bibinfo
  {volume} {120}},\ \bibinfo {pages} {152501} (\bibinfo {year}
  {2018})}\BibitemShut {NoStop}%
\bibitem [{\citenamefont {Kaji}\ \emph {et~al.}(2008)\citenamefont {Kaji},
  \citenamefont {Morimoto}, \citenamefont {Yoneda}, \citenamefont {Hasebe},
  \citenamefont {Yoshida}, \citenamefont {Haba}, \citenamefont {Goto},
  \citenamefont {Kudo},\ and\ \citenamefont {Morita}}]{KAJI2008198}%
  \BibitemOpen
  \bibfield  {author} {\bibinfo {author} {\bibfnamefont {D.}~\bibnamefont
  {Kaji}}, \bibinfo {author} {\bibfnamefont {K.}~\bibnamefont {Morimoto}},
  \bibinfo {author} {\bibfnamefont {A.}~\bibnamefont {Yoneda}}, \bibinfo
  {author} {\bibfnamefont {H.}~\bibnamefont {Hasebe}}, \bibinfo {author}
  {\bibfnamefont {A.}~\bibnamefont {Yoshida}}, \bibinfo {author} {\bibfnamefont
  {H.}~\bibnamefont {Haba}}, \bibinfo {author} {\bibfnamefont {S.}~\bibnamefont
  {Goto}}, \bibinfo {author} {\bibfnamefont {H.}~\bibnamefont {Kudo}}, \ and\
  \bibinfo {author} {\bibfnamefont {K.}~\bibnamefont {Morita}},\ }\href
  {\doibase https://doi.org/10.1016/j.nima.2008.02.090} {\bibfield  {journal}
  {\bibinfo  {journal} {Nucl. Instrum. Meth. A}\ }\textbf {\bibinfo {volume}
  {590}},\ \bibinfo {pages} {198 } (\bibinfo {year} {2008})}\BibitemShut
  {NoStop}%
\bibitem [{\citenamefont {Odera}\ \emph {et~al.}(1984)\citenamefont {Odera},
  \citenamefont {Chiba}, \citenamefont {Tonuma}, \citenamefont {Hemmi},
  \citenamefont {Miyazawa}, \citenamefont {Inoue}, \citenamefont {Kambara},
  \citenamefont {Kase}, \citenamefont {Kubo},\ and\ \citenamefont
  {Yoshida}}]{RILAC1984}%
  \BibitemOpen
  \bibfield  {author} {\bibinfo {author} {\bibfnamefont {M.}~\bibnamefont
  {Odera}}, \bibinfo {author} {\bibfnamefont {Y.}~\bibnamefont {Chiba}},
  \bibinfo {author} {\bibfnamefont {T.}~\bibnamefont {Tonuma}}, \bibinfo
  {author} {\bibfnamefont {M.}~\bibnamefont {Hemmi}}, \bibinfo {author}
  {\bibfnamefont {Y.}~\bibnamefont {Miyazawa}}, \bibinfo {author}
  {\bibfnamefont {T.}~\bibnamefont {Inoue}}, \bibinfo {author} {\bibfnamefont
  {T.}~\bibnamefont {Kambara}}, \bibinfo {author} {\bibfnamefont
  {M.}~\bibnamefont {Kase}}, \bibinfo {author} {\bibfnamefont {T.}~\bibnamefont
  {Kubo}}, \ and\ \bibinfo {author} {\bibfnamefont {F.}~\bibnamefont
  {Yoshida}},\ }\href {\doibase https://doi.org/10.1016/0168-9002(84)90121-9}
  {\bibfield  {journal} {\bibinfo  {journal} {Nucl. Instr. Meth. B}\ }\textbf
  {\bibinfo {volume} {227}},\ \bibinfo {pages} {187 } (\bibinfo {year}
  {1984})}\BibitemShut {NoStop}%
\bibitem [{\citenamefont {Kaji}\ \emph {et~al.}(2013)\citenamefont {Kaji},
  \citenamefont {Morimoto}, \citenamefont {Sato}, \citenamefont {Yoneda},\ and\
  \citenamefont {Morita}}]{KAJI2013311}%
  \BibitemOpen
  \bibfield  {author} {\bibinfo {author} {\bibfnamefont {D.}~\bibnamefont
  {Kaji}}, \bibinfo {author} {\bibfnamefont {K.}~\bibnamefont {Morimoto}},
  \bibinfo {author} {\bibfnamefont {N.}~\bibnamefont {Sato}}, \bibinfo {author}
  {\bibfnamefont {A.}~\bibnamefont {Yoneda}}, \ and\ \bibinfo {author}
  {\bibfnamefont {K.}~\bibnamefont {Morita}},\ }\href {\doibase
  http://dx.doi.org/10.1016/j.nimb.2013.05.085} {\bibfield  {journal} {\bibinfo
   {journal} {Nucl. Instr. Meth. B}\ }\textbf {\bibinfo {volume} {317}},\
  \bibinfo {pages} {311 } (\bibinfo {year} {2013})}\BibitemShut {NoStop}%
\bibitem [{\citenamefont {Schury}\ \emph
  {et~al.}(2017{\natexlab{b}})\citenamefont {Schury}, \citenamefont {Wada},
  \citenamefont {Ito}, \citenamefont {Kaji}, \citenamefont {Haba},
  \citenamefont {Hirayama}, \citenamefont {Kimura}, \citenamefont {Koura},
  \citenamefont {MacCormick}, \citenamefont {Miyatake}, \citenamefont {Moon},
  \citenamefont {Morimoto}, \citenamefont {Morita}, \citenamefont {Murray},
  \citenamefont {Ozawa}, \citenamefont {Rosenbusch}, \citenamefont {Reponen},
  \citenamefont {Takamine}, \citenamefont {Tanaka}, \citenamefont {Watanabe},\
  and\ \citenamefont {Wollnik}}]{SCHURY2017160}%
  \BibitemOpen
  \bibfield  {author} {\bibinfo {author} {\bibfnamefont {P.}~\bibnamefont
  {Schury}}, \bibinfo {author} {\bibfnamefont {M.}~\bibnamefont {Wada}},
  \bibinfo {author} {\bibfnamefont {Y.}~\bibnamefont {Ito}}, \bibinfo {author}
  {\bibfnamefont {D.}~\bibnamefont {Kaji}}, \bibinfo {author} {\bibfnamefont
  {H.}~\bibnamefont {Haba}}, \bibinfo {author} {\bibfnamefont {Y.}~\bibnamefont
  {Hirayama}}, \bibinfo {author} {\bibfnamefont {S.}~\bibnamefont {Kimura}},
  \bibinfo {author} {\bibfnamefont {H.}~\bibnamefont {Koura}}, \bibinfo
  {author} {\bibfnamefont {M.}~\bibnamefont {MacCormick}}, \bibinfo {author}
  {\bibfnamefont {H.}~\bibnamefont {Miyatake}}, \bibinfo {author}
  {\bibfnamefont {J.}~\bibnamefont {Moon}}, \bibinfo {author} {\bibfnamefont
  {K.}~\bibnamefont {Morimoto}}, \bibinfo {author} {\bibfnamefont
  {K.}~\bibnamefont {Morita}}, \bibinfo {author} {\bibfnamefont
  {I.}~\bibnamefont {Murray}}, \bibinfo {author} {\bibfnamefont
  {A.}~\bibnamefont {Ozawa}}, \bibinfo {author} {\bibfnamefont
  {M.}~\bibnamefont {Rosenbusch}}, \bibinfo {author} {\bibfnamefont
  {M.}~\bibnamefont {Reponen}}, \bibinfo {author} {\bibfnamefont
  {A.}~\bibnamefont {Takamine}}, \bibinfo {author} {\bibfnamefont
  {T.}~\bibnamefont {Tanaka}}, \bibinfo {author} {\bibfnamefont
  {Y.}~\bibnamefont {Watanabe}}, \ and\ \bibinfo {author} {\bibfnamefont
  {H.}~\bibnamefont {Wollnik}},\ }\href {\doibase
  https://doi.org/10.1016/j.nimb.2017.06.014} {\bibfield  {journal} {\bibinfo
  {journal} {Nucl. Instr. Meth. B}\ }\textbf {\bibinfo {volume} {407}},\
  \bibinfo {pages} {160 } (\bibinfo {year} {2017}{\natexlab{b}})}\BibitemShut
  {NoStop}%
\bibitem [{\citenamefont {Wada}\ \emph {et~al.}(2003)\citenamefont {Wada},
  \citenamefont {Ishida}, \citenamefont {Nakamura}, \citenamefont {Yamazaki},
  \citenamefont {Kambara}, \citenamefont {Ohyama}, \citenamefont {Kanai},
  \citenamefont {Kojima}, \citenamefont {Nakai}, \citenamefont {Ohshima},
  \citenamefont {Yoshida}, \citenamefont {Kubo}, \citenamefont {Matsuo},
  \citenamefont {Fukuyama}, \citenamefont {Okada}, \citenamefont {Sonoda},
  \citenamefont {Ohtani}, \citenamefont {Noda}, \citenamefont {Kawakami},\ and\
  \citenamefont {Katayama}}]{WADA2003570}%
  \BibitemOpen
  \bibfield  {author} {\bibinfo {author} {\bibfnamefont {M.}~\bibnamefont
  {Wada}}, \bibinfo {author} {\bibfnamefont {Y.}~\bibnamefont {Ishida}},
  \bibinfo {author} {\bibfnamefont {T.}~\bibnamefont {Nakamura}}, \bibinfo
  {author} {\bibfnamefont {Y.}~\bibnamefont {Yamazaki}}, \bibinfo {author}
  {\bibfnamefont {T.}~\bibnamefont {Kambara}}, \bibinfo {author} {\bibfnamefont
  {H.}~\bibnamefont {Ohyama}}, \bibinfo {author} {\bibfnamefont
  {Y.}~\bibnamefont {Kanai}}, \bibinfo {author} {\bibfnamefont {T.~M.}\
  \bibnamefont {Kojima}}, \bibinfo {author} {\bibfnamefont {Y.}~\bibnamefont
  {Nakai}}, \bibinfo {author} {\bibfnamefont {N.}~\bibnamefont {Ohshima}},
  \bibinfo {author} {\bibfnamefont {A.}~\bibnamefont {Yoshida}}, \bibinfo
  {author} {\bibfnamefont {T.}~\bibnamefont {Kubo}}, \bibinfo {author}
  {\bibfnamefont {Y.}~\bibnamefont {Matsuo}}, \bibinfo {author} {\bibfnamefont
  {Y.}~\bibnamefont {Fukuyama}}, \bibinfo {author} {\bibfnamefont
  {K.}~\bibnamefont {Okada}}, \bibinfo {author} {\bibfnamefont
  {T.}~\bibnamefont {Sonoda}}, \bibinfo {author} {\bibfnamefont
  {S.}~\bibnamefont {Ohtani}}, \bibinfo {author} {\bibfnamefont
  {K.}~\bibnamefont {Noda}}, \bibinfo {author} {\bibfnamefont {H.}~\bibnamefont
  {Kawakami}}, \ and\ \bibinfo {author} {\bibfnamefont {I.}~\bibnamefont
  {Katayama}},\ }\href {\doibase https://doi.org/10.1016/S0168-583X(02)02151-1}
  {\bibfield  {journal} {\bibinfo  {journal} {Nucl. Instrum. Meth. B}\ }\textbf
  {\bibinfo {volume} {204}},\ \bibinfo {pages} {570 } (\bibinfo {year}
  {2003})}\BibitemShut {NoStop}%
\bibitem [{\citenamefont {Bollen}(2011)}]{BOLLEN2011131}%
  \BibitemOpen
  \bibfield  {author} {\bibinfo {author} {\bibfnamefont {G.}~\bibnamefont
  {Bollen}},\ }\href {\doibase https://doi.org/10.1016/j.ijms.2010.09.032}
  {\bibfield  {journal} {\bibinfo  {journal} {Int. J. Mass Spectrom.}\ }\textbf
  {\bibinfo {volume} {299}},\ \bibinfo {pages} {131 } (\bibinfo {year}
  {2011})}\BibitemShut {NoStop}%
\bibitem [{\citenamefont {Arai}\ \emph {et~al.}(2014)\citenamefont {Arai},
  \citenamefont {Ito}, \citenamefont {Wada}, \citenamefont {Schury},
  \citenamefont {Sonoda},\ and\ \citenamefont {Mita}}]{ARAI201456}%
  \BibitemOpen
  \bibfield  {author} {\bibinfo {author} {\bibfnamefont {F.}~\bibnamefont
  {Arai}}, \bibinfo {author} {\bibfnamefont {Y.}~\bibnamefont {Ito}}, \bibinfo
  {author} {\bibfnamefont {M.}~\bibnamefont {Wada}}, \bibinfo {author}
  {\bibfnamefont {P.}~\bibnamefont {Schury}}, \bibinfo {author} {\bibfnamefont
  {T.}~\bibnamefont {Sonoda}}, \ and\ \bibinfo {author} {\bibfnamefont
  {H.}~\bibnamefont {Mita}},\ }\href {\doibase
  https://doi.org/10.1016/j.ijms.2014.01.005} {\bibfield  {journal} {\bibinfo
  {journal} {Int. J. Mass Spectrom.}\ }\textbf {\bibinfo {volume} {362}},\
  \bibinfo {pages} {56 } (\bibinfo {year} {2014})}\BibitemShut {NoStop}%
\bibitem [{\citenamefont {Xu}\ \emph {et~al.}(1993)\citenamefont {Xu},
  \citenamefont {Wada}, \citenamefont {Tanaka}, \citenamefont {Kawakami},
  \citenamefont {Katayama},\ and\ \citenamefont {Ohtani}}]{XU1993274}%
  \BibitemOpen
  \bibfield  {author} {\bibinfo {author} {\bibfnamefont {H.~J.}\ \bibnamefont
  {Xu}}, \bibinfo {author} {\bibfnamefont {M.}~\bibnamefont {Wada}}, \bibinfo
  {author} {\bibfnamefont {J.}~\bibnamefont {Tanaka}}, \bibinfo {author}
  {\bibfnamefont {H.}~\bibnamefont {Kawakami}}, \bibinfo {author}
  {\bibfnamefont {I.}~\bibnamefont {Katayama}}, \ and\ \bibinfo {author}
  {\bibfnamefont {S.}~\bibnamefont {Ohtani}},\ }\href {\doibase
  https://doi.org/10.1016/0168-9002(93)91166-K} {\bibfield  {journal} {\bibinfo
   {journal} {Nucl. Instrum. Meth. A}\ }\textbf {\bibinfo {volume} {333}},\
  \bibinfo {pages} {274 } (\bibinfo {year} {1993})}\BibitemShut {NoStop}%
\bibitem [{\citenamefont {Schury}\ \emph {et~al.}(2009)\citenamefont {Schury},
  \citenamefont {Okada}, \citenamefont {Shchepunov}, \citenamefont {Sonoda},
  \citenamefont {Takamine}, \citenamefont {Wada}, \citenamefont {Wollnik},\
  and\ \citenamefont {Yamazaki}}]{Schury2009}%
  \BibitemOpen
  \bibfield  {author} {\bibinfo {author} {\bibfnamefont {P.}~\bibnamefont
  {Schury}}, \bibinfo {author} {\bibfnamefont {K.}~\bibnamefont {Okada}},
  \bibinfo {author} {\bibfnamefont {S.}~\bibnamefont {Shchepunov}}, \bibinfo
  {author} {\bibfnamefont {T.}~\bibnamefont {Sonoda}}, \bibinfo {author}
  {\bibfnamefont {A.}~\bibnamefont {Takamine}}, \bibinfo {author}
  {\bibfnamefont {M.}~\bibnamefont {Wada}}, \bibinfo {author} {\bibfnamefont
  {H.}~\bibnamefont {Wollnik}}, \ and\ \bibinfo {author} {\bibfnamefont
  {Y.}~\bibnamefont {Yamazaki}},\ }\href {\doibase 10.1140/epja/i2009-10882-6}
  {\bibfield  {journal} {\bibinfo  {journal} {Eur. Phys. J. A}\ }\textbf
  {\bibinfo {volume} {42}},\ \bibinfo {pages} {343} (\bibinfo {year}
  {2009})}\BibitemShut {NoStop}%
\bibitem [{\citenamefont {Bradbury}\ and\ \citenamefont
  {Nielsen}(1936)}]{Bradbury1936}%
  \BibitemOpen
  \bibfield  {author} {\bibinfo {author} {\bibfnamefont {N.~E.}\ \bibnamefont
  {Bradbury}}\ and\ \bibinfo {author} {\bibfnamefont {R.~A.}\ \bibnamefont
  {Nielsen}},\ }\href {\doibase 10.1103/PhysRev.49.388} {\bibfield  {journal}
  {\bibinfo  {journal} {Phys. Rev.}\ }\textbf {\bibinfo {volume} {49}},\
  \bibinfo {pages} {388} (\bibinfo {year} {1936})}\BibitemShut {NoStop}%
\bibitem [{\citenamefont {Brun}\ and\ \citenamefont
  {Rademakers}(1997)}]{BRUN199781}%
  \BibitemOpen
  \bibfield  {author} {\bibinfo {author} {\bibfnamefont {R.}~\bibnamefont
  {Brun}}\ and\ \bibinfo {author} {\bibfnamefont {F.}~\bibnamefont
  {Rademakers}},\ }\href {\doibase
  https://doi.org/10.1016/S0168-9002(97)00048-X} {\bibfield  {journal}
  {\bibinfo  {journal} {Nucl. Instr. Meth. B}\ }\textbf {\bibinfo {volume}
  {389}},\ \bibinfo {pages} {81 } (\bibinfo {year} {1997})}\BibitemShut
  {NoStop}%
\bibitem [{\citenamefont {Koskelo}\ \emph {et~al.}(1981)\citenamefont
  {Koskelo}, \citenamefont {Aarnio},\ and\ \citenamefont
  {Routti}}]{KOSKELO198111}%
  \BibitemOpen
  \bibfield  {author} {\bibinfo {author} {\bibfnamefont {M.~J.}\ \bibnamefont
  {Koskelo}}, \bibinfo {author} {\bibfnamefont {P.~A.}\ \bibnamefont {Aarnio}},
  \ and\ \bibinfo {author} {\bibfnamefont {J.~T.}\ \bibnamefont {Routti}},\
  }\href {\doibase https://doi.org/10.1016/0010-4655(81)90104-1} {\bibfield
  {journal} {\bibinfo  {journal} {Comput. Phys. Commun.}\ }\textbf {\bibinfo
  {volume} {24}},\ \bibinfo {pages} {11 } (\bibinfo {year} {1981})}\BibitemShut
  {NoStop}%
\bibitem [{\citenamefont {Lan}\ and\ \citenamefont
  {Jorgenson}(2001)}]{LAN20011}%
  \BibitemOpen
  \bibfield  {author} {\bibinfo {author} {\bibfnamefont {K.}~\bibnamefont
  {Lan}}\ and\ \bibinfo {author} {\bibfnamefont {J.~W.}\ \bibnamefont
  {Jorgenson}},\ }\href {\doibase
  https://doi.org/10.1016/S0021-9673(01)00594-5} {\bibfield  {journal}
  {\bibinfo  {journal} {J. Chromatogr. A}\ }\textbf {\bibinfo {volume} {915}},\
  \bibinfo {pages} {1 } (\bibinfo {year} {2001})}\BibitemShut {NoStop}%
\bibitem [{\citenamefont {Ito}\ \emph {et~al.}(2013)\citenamefont {Ito},
  \citenamefont {Schury}, \citenamefont {Wada}, \citenamefont {Naimi},
  \citenamefont {Sonoda}, \citenamefont {Mita}, \citenamefont {Arai},
  \citenamefont {Takamine}, \citenamefont {Okada}, \citenamefont {Ozawa},\ and\
  \citenamefont {Wollnik}}]{ITO2013}%
  \BibitemOpen
  \bibfield  {author} {\bibinfo {author} {\bibfnamefont {Y.}~\bibnamefont
  {Ito}}, \bibinfo {author} {\bibfnamefont {P.}~\bibnamefont {Schury}},
  \bibinfo {author} {\bibfnamefont {M.}~\bibnamefont {Wada}}, \bibinfo {author}
  {\bibfnamefont {S.}~\bibnamefont {Naimi}}, \bibinfo {author} {\bibfnamefont
  {T.}~\bibnamefont {Sonoda}}, \bibinfo {author} {\bibfnamefont
  {H.}~\bibnamefont {Mita}}, \bibinfo {author} {\bibfnamefont {F.}~\bibnamefont
  {Arai}}, \bibinfo {author} {\bibfnamefont {A.}~\bibnamefont {Takamine}},
  \bibinfo {author} {\bibfnamefont {K.}~\bibnamefont {Okada}}, \bibinfo
  {author} {\bibfnamefont {A.}~\bibnamefont {Ozawa}}, \ and\ \bibinfo {author}
  {\bibfnamefont {H.}~\bibnamefont {Wollnik}},\ }\href {\doibase
  10.1103/PhysRevC.88.011306} {\bibfield  {journal} {\bibinfo  {journal} {Phys.
  Rev. C}\ }\textbf {\bibinfo {volume} {88}},\ \bibinfo {pages} {011306}
  (\bibinfo {year} {2013})}\BibitemShut {NoStop}%
\bibitem [{\citenamefont {Wienholtz}\ \emph {et~al.}(2013)\citenamefont
  {Wienholtz}, \citenamefont {Beck}, \citenamefont {Blaum}, \citenamefont
  {Borgmann}, \citenamefont {Breitenfeldt}, \citenamefont {Cakirli},
  \citenamefont {George}, \citenamefont {Herfurth}, \citenamefont {Holt},
  \citenamefont {Kowalska}, \citenamefont {Kreim}, \citenamefont {Lunney},
  \citenamefont {Manea}, \citenamefont {Men\`endez}, \citenamefont {Neidherr},
  \citenamefont {Rosenbusch}, \citenamefont {Schweikhard}, \citenamefont
  {Schwenk}, \citenamefont {Simonis}, \citenamefont {Stanja}, \citenamefont
  {Wolf},\ and\ \citenamefont {Zuber}}]{Wienholtz2013}%
  \BibitemOpen
  \bibfield  {author} {\bibinfo {author} {\bibfnamefont {F.}~\bibnamefont
  {Wienholtz}}, \bibinfo {author} {\bibfnamefont {D.}~\bibnamefont {Beck}},
  \bibinfo {author} {\bibfnamefont {K.}~\bibnamefont {Blaum}}, \bibinfo
  {author} {\bibfnamefont {C.}~\bibnamefont {Borgmann}}, \bibinfo {author}
  {\bibfnamefont {M.}~\bibnamefont {Breitenfeldt}}, \bibinfo {author}
  {\bibfnamefont {R.~B.}\ \bibnamefont {Cakirli}}, \bibinfo {author}
  {\bibfnamefont {S.}~\bibnamefont {George}}, \bibinfo {author} {\bibfnamefont
  {F.}~\bibnamefont {Herfurth}}, \bibinfo {author} {\bibfnamefont {J.~D.}\
  \bibnamefont {Holt}}, \bibinfo {author} {\bibfnamefont {M.}~\bibnamefont
  {Kowalska}}, \bibinfo {author} {\bibfnamefont {S.}~\bibnamefont {Kreim}},
  \bibinfo {author} {\bibfnamefont {D.}~\bibnamefont {Lunney}}, \bibinfo
  {author} {\bibfnamefont {V.}~\bibnamefont {Manea}}, \bibinfo {author}
  {\bibfnamefont {J.}~\bibnamefont {Men\`endez}}, \bibinfo {author}
  {\bibfnamefont {D.}~\bibnamefont {Neidherr}}, \bibinfo {author}
  {\bibfnamefont {M.}~\bibnamefont {Rosenbusch}}, \bibinfo {author}
  {\bibfnamefont {L.}~\bibnamefont {Schweikhard}}, \bibinfo {author}
  {\bibfnamefont {A.}~\bibnamefont {Schwenk}}, \bibinfo {author} {\bibfnamefont
  {J.}~\bibnamefont {Simonis}}, \bibinfo {author} {\bibfnamefont
  {J.}~\bibnamefont {Stanja}}, \bibinfo {author} {\bibfnamefont {R.~N.}\
  \bibnamefont {Wolf}}, \ and\ \bibinfo {author} {\bibfnamefont
  {K.}~\bibnamefont {Zuber}},\ }\href {http://dx.doi.org/10.1038/nature12226}
  {\bibfield  {journal} {\bibinfo  {journal} {Nature}\ }\textbf {\bibinfo
  {volume} {498}},\ \bibinfo {pages} {346} (\bibinfo {year}
  {2013})}\BibitemShut {NoStop}%
\bibitem [{\citenamefont {Droese}\ \emph {et~al.}(2013)\citenamefont {Droese},
  \citenamefont {Ackermann}, \citenamefont {Andersson}, \citenamefont {Blaum},
  \citenamefont {Block}, \citenamefont {Dworschak}, \citenamefont {Eibach},
  \citenamefont {Eliseev}, \citenamefont {Forsberg}, \citenamefont {Haettner},
  \citenamefont {Herfurth}, \citenamefont {He{\ss}berger}, \citenamefont
  {Hofmann}, \citenamefont {Ketelaer}, \citenamefont {Marx}, \citenamefont
  {Minaya~Ramirez}, \citenamefont {Nesterenko}, \citenamefont {Novikov},
  \citenamefont {Pla{\ss}}, \citenamefont {Rodr{\'i}guez}, \citenamefont
  {Rudolph}, \citenamefont {Scheidenberger}, \citenamefont {Schweikhard},
  \citenamefont {Stolze}, \citenamefont {Thirolf},\ and\ \citenamefont
  {Weber}}]{Droese2013}%
  \BibitemOpen
  \bibfield  {author} {\bibinfo {author} {\bibfnamefont {C.}~\bibnamefont
  {Droese}}, \bibinfo {author} {\bibfnamefont {D.}~\bibnamefont {Ackermann}},
  \bibinfo {author} {\bibfnamefont {L.~L.}\ \bibnamefont {Andersson}}, \bibinfo
  {author} {\bibfnamefont {K.}~\bibnamefont {Blaum}}, \bibinfo {author}
  {\bibfnamefont {M.}~\bibnamefont {Block}}, \bibinfo {author} {\bibfnamefont
  {M.}~\bibnamefont {Dworschak}}, \bibinfo {author} {\bibfnamefont
  {M.}~\bibnamefont {Eibach}}, \bibinfo {author} {\bibfnamefont
  {S.}~\bibnamefont {Eliseev}}, \bibinfo {author} {\bibfnamefont
  {U.}~\bibnamefont {Forsberg}}, \bibinfo {author} {\bibfnamefont
  {E.}~\bibnamefont {Haettner}}, \bibinfo {author} {\bibfnamefont
  {F.}~\bibnamefont {Herfurth}}, \bibinfo {author} {\bibfnamefont {F.~P.}\
  \bibnamefont {He{\ss}berger}}, \bibinfo {author} {\bibfnamefont
  {S.}~\bibnamefont {Hofmann}}, \bibinfo {author} {\bibfnamefont
  {J.}~\bibnamefont {Ketelaer}}, \bibinfo {author} {\bibfnamefont
  {G.}~\bibnamefont {Marx}}, \bibinfo {author} {\bibfnamefont {E.}~\bibnamefont
  {Minaya~Ramirez}}, \bibinfo {author} {\bibfnamefont {D.}~\bibnamefont
  {Nesterenko}}, \bibinfo {author} {\bibfnamefont {Y.~N.}\ \bibnamefont
  {Novikov}}, \bibinfo {author} {\bibfnamefont {W.~R.}\ \bibnamefont
  {Pla{\ss}}}, \bibinfo {author} {\bibfnamefont {D.}~\bibnamefont
  {Rodr{\'i}guez}}, \bibinfo {author} {\bibfnamefont {D.}~\bibnamefont
  {Rudolph}}, \bibinfo {author} {\bibfnamefont {C.}~\bibnamefont
  {Scheidenberger}}, \bibinfo {author} {\bibfnamefont {L.}~\bibnamefont
  {Schweikhard}}, \bibinfo {author} {\bibfnamefont {S.}~\bibnamefont {Stolze}},
  \bibinfo {author} {\bibfnamefont {P.~G.}\ \bibnamefont {Thirolf}}, \ and\
  \bibinfo {author} {\bibfnamefont {C.}~\bibnamefont {Weber}},\ }\href
  {\doibase 10.1140/epja/i2013-13013-0} {\bibfield  {journal} {\bibinfo
  {journal} {Eur. Phys. J. A}\ }\textbf {\bibinfo {volume} {49}},\ \bibinfo
  {pages} {13} (\bibinfo {year} {2013})}\BibitemShut {NoStop}%
\bibitem [{\citenamefont {He{\ss}berger}\ \emph {et~al.}(2000)\citenamefont
  {He{\ss}berger}, \citenamefont {Hofmann}, \citenamefont {Ackermann},
  \citenamefont {Ninov}, \citenamefont {Leino}, \citenamefont {Saro},
  \citenamefont {Andreyev}, \citenamefont {Lavrentev}, \citenamefont {Popeko},\
  and\ \citenamefont {Yeremin}}]{2000He17}%
  \BibitemOpen
  \bibfield  {author} {\bibinfo {author} {\bibfnamefont {F.~P.}\ \bibnamefont
  {He{\ss}berger}}, \bibinfo {author} {\bibfnamefont {S.}~\bibnamefont
  {Hofmann}}, \bibinfo {author} {\bibfnamefont {D.}~\bibnamefont {Ackermann}},
  \bibinfo {author} {\bibfnamefont {V.}~\bibnamefont {Ninov}}, \bibinfo
  {author} {\bibfnamefont {M.}~\bibnamefont {Leino}}, \bibinfo {author}
  {\bibfnamefont {S.}~\bibnamefont {Saro}}, \bibinfo {author} {\bibfnamefont
  {A.}~\bibnamefont {Andreyev}}, \bibinfo {author} {\bibfnamefont
  {A.}~\bibnamefont {Lavrentev}}, \bibinfo {author} {\bibfnamefont {A.~G.}\
  \bibnamefont {Popeko}}, \ and\ \bibinfo {author} {\bibfnamefont {A.~V.}\
  \bibnamefont {Yeremin}},\ }\href {\doibase 10.1007/s100500070075} {\bibfield
  {journal} {\bibinfo  {journal} {Eur. Phys. J. A}\ }\textbf {\bibinfo {volume}
  {8}},\ \bibinfo {pages} {521} (\bibinfo {year} {2000})}\BibitemShut {NoStop}%
\bibitem [{\citenamefont {Kowalska}\ \emph {et~al.}(2009)\citenamefont
  {Kowalska}, \citenamefont {Naimi}, \citenamefont {Agramunt}, \citenamefont
  {Algora}, \citenamefont {Audi}, \citenamefont {Beck}, \citenamefont {Blank},
  \citenamefont {Blaum}, \citenamefont {B{\"o}hm}, \citenamefont
  {Breitenfeldt}, \citenamefont {Estevez}, \citenamefont {Fraile},
  \citenamefont {George}, \citenamefont {Herfurth}, \citenamefont {Herlert},
  \citenamefont {Kellerbauer}, \citenamefont {Lunney}, \citenamefont
  {Minaya-Ramirez}, \citenamefont {Neidherr}, \citenamefont {Olaizola},
  \citenamefont {Riisager}, \citenamefont {Rosenbusch}, \citenamefont {Rubio},
  \citenamefont {Schwarz}, \citenamefont {Schweikhard},\ and\ \citenamefont
  {Warring}}]{Kowalska2009}%
  \BibitemOpen
  \bibfield  {author} {\bibinfo {author} {\bibfnamefont {M.}~\bibnamefont
  {Kowalska}}, \bibinfo {author} {\bibfnamefont {S.}~\bibnamefont {Naimi}},
  \bibinfo {author} {\bibfnamefont {J.}~\bibnamefont {Agramunt}}, \bibinfo
  {author} {\bibfnamefont {A.}~\bibnamefont {Algora}}, \bibinfo {author}
  {\bibfnamefont {G.}~\bibnamefont {Audi}}, \bibinfo {author} {\bibfnamefont
  {D.}~\bibnamefont {Beck}}, \bibinfo {author} {\bibfnamefont {B.}~\bibnamefont
  {Blank}}, \bibinfo {author} {\bibfnamefont {K.}~\bibnamefont {Blaum}},
  \bibinfo {author} {\bibfnamefont {C.}~\bibnamefont {B{\"o}hm}}, \bibinfo
  {author} {\bibfnamefont {M.}~\bibnamefont {Breitenfeldt}}, \bibinfo {author}
  {\bibfnamefont {E.}~\bibnamefont {Estevez}}, \bibinfo {author} {\bibfnamefont
  {L.~M.}\ \bibnamefont {Fraile}}, \bibinfo {author} {\bibfnamefont
  {S.}~\bibnamefont {George}}, \bibinfo {author} {\bibfnamefont
  {F.}~\bibnamefont {Herfurth}}, \bibinfo {author} {\bibfnamefont
  {A.}~\bibnamefont {Herlert}}, \bibinfo {author} {\bibfnamefont
  {A.}~\bibnamefont {Kellerbauer}}, \bibinfo {author} {\bibfnamefont
  {D.}~\bibnamefont {Lunney}}, \bibinfo {author} {\bibfnamefont
  {E.}~\bibnamefont {Minaya-Ramirez}}, \bibinfo {author} {\bibfnamefont
  {D.}~\bibnamefont {Neidherr}}, \bibinfo {author} {\bibfnamefont
  {B.}~\bibnamefont {Olaizola}}, \bibinfo {author} {\bibfnamefont
  {K.}~\bibnamefont {Riisager}}, \bibinfo {author} {\bibfnamefont
  {M.}~\bibnamefont {Rosenbusch}}, \bibinfo {author} {\bibfnamefont
  {B.}~\bibnamefont {Rubio}}, \bibinfo {author} {\bibfnamefont
  {S.}~\bibnamefont {Schwarz}}, \bibinfo {author} {\bibfnamefont
  {L.}~\bibnamefont {Schweikhard}}, \ and\ \bibinfo {author} {\bibfnamefont
  {U.}~\bibnamefont {Warring}},\ }\href {\doibase 10.1140/epja/i2009-10835-1}
  {\bibfield  {journal} {\bibinfo  {journal} {Eur. Phys. J. A}\ }\textbf
  {\bibinfo {volume} {42}},\ \bibinfo {pages} {351} (\bibinfo {year}
  {2009})}\BibitemShut {NoStop}%
\bibitem [{\citenamefont {B\"ohm}\ \emph {et~al.}(2014)\citenamefont {B\"ohm},
  \citenamefont {Borgmann}, \citenamefont {Audi}, \citenamefont {Beck},
  \citenamefont {Blaum}, \citenamefont {Breitenfeldt}, \citenamefont {Cakirli},
  \citenamefont {Cocolios}, \citenamefont {Eliseev}, \citenamefont {George},
  \citenamefont {Herfurth}, \citenamefont {Herlert}, \citenamefont {Kowalska},
  \citenamefont {Kreim}, \citenamefont {Lunney}, \citenamefont {Manea},
  \citenamefont {Minaya~Ramirez}, \citenamefont {Naimi}, \citenamefont
  {Neidherr}, \citenamefont {Rosenbusch}, \citenamefont {Schweikhard},
  \citenamefont {Stanja}, \citenamefont {Wang}, \citenamefont {Wolf},\ and\
  \citenamefont {Zuber}}]{2014Bo26}%
  \BibitemOpen
  \bibfield  {author} {\bibinfo {author} {\bibfnamefont {C.}~\bibnamefont
  {B\"ohm}}, \bibinfo {author} {\bibfnamefont {C.}~\bibnamefont {Borgmann}},
  \bibinfo {author} {\bibfnamefont {G.}~\bibnamefont {Audi}}, \bibinfo {author}
  {\bibfnamefont {D.}~\bibnamefont {Beck}}, \bibinfo {author} {\bibfnamefont
  {K.}~\bibnamefont {Blaum}}, \bibinfo {author} {\bibfnamefont
  {M.}~\bibnamefont {Breitenfeldt}}, \bibinfo {author} {\bibfnamefont {R.~B.}\
  \bibnamefont {Cakirli}}, \bibinfo {author} {\bibfnamefont {T.~E.}\
  \bibnamefont {Cocolios}}, \bibinfo {author} {\bibfnamefont {S.}~\bibnamefont
  {Eliseev}}, \bibinfo {author} {\bibfnamefont {S.}~\bibnamefont {George}},
  \bibinfo {author} {\bibfnamefont {F.}~\bibnamefont {Herfurth}}, \bibinfo
  {author} {\bibfnamefont {A.}~\bibnamefont {Herlert}}, \bibinfo {author}
  {\bibfnamefont {M.}~\bibnamefont {Kowalska}}, \bibinfo {author}
  {\bibfnamefont {S.}~\bibnamefont {Kreim}}, \bibinfo {author} {\bibfnamefont
  {D.}~\bibnamefont {Lunney}}, \bibinfo {author} {\bibfnamefont
  {V.}~\bibnamefont {Manea}}, \bibinfo {author} {\bibfnamefont
  {E.}~\bibnamefont {Minaya~Ramirez}}, \bibinfo {author} {\bibfnamefont
  {S.}~\bibnamefont {Naimi}}, \bibinfo {author} {\bibfnamefont
  {D.}~\bibnamefont {Neidherr}}, \bibinfo {author} {\bibfnamefont
  {M.}~\bibnamefont {Rosenbusch}}, \bibinfo {author} {\bibfnamefont
  {L.}~\bibnamefont {Schweikhard}}, \bibinfo {author} {\bibfnamefont
  {J.}~\bibnamefont {Stanja}}, \bibinfo {author} {\bibfnamefont
  {M.}~\bibnamefont {Wang}}, \bibinfo {author} {\bibfnamefont {R.~N.}\
  \bibnamefont {Wolf}}, \ and\ \bibinfo {author} {\bibfnamefont
  {K.}~\bibnamefont {Zuber}},\ }\href {\doibase 10.1103/PhysRevC.90.044307}
  {\bibfield  {journal} {\bibinfo  {journal} {Phys. Rev. C}\ }\textbf {\bibinfo
  {volume} {90}},\ \bibinfo {pages} {044307} (\bibinfo {year}
  {2014})}\BibitemShut {NoStop}%
\end{thebibliography}%

\end{document}